\documentclass[aps,prl,superscriptaddress,amsfonts,amsmath,amssymb,showpacs,floatfix,preprint,longbibliography]{revtex4-2}

\usepackage{url}
\usepackage{bm}
\usepackage{graphicx}
\usepackage{amsmath}
\usepackage{amstext}
\usepackage{amssymb}
\usepackage{amsfonts}
\usepackage{amsbsy}
\usepackage{verbatim}
\usepackage{xcolor}
\usepackage[colorlinks=true, urlcolor=blue, linkcolor=blue, citecolor=blue, pdftex]{hyperref}
\usepackage{multirow}
\usepackage{floatrow}
\usepackage{float}
\usepackage{gensymb}
\usepackage{textcomp}
\usepackage{enumitem}
\usepackage[version=4]{mhchem}
\usepackage{afterpage}
\usepackage{pbox}
\usepackage{makecell}
\usepackage{array,booktabs}
\usepackage{dsfont}
\usepackage{siunitx}
\usepackage[utf8]{inputenc}
\usepackage{braket}
\usepackage[normalem]{ulem}

\newcommand{\kni}{K$_2$Ni$_2$(SO$_4$)$_3$}
\newcommand{\kmg}{K$_2$Mg$_2$(SO$_4$)$_3$}

\begin{document}


\title{Magnetic field induced quantum spin liquid in the two coupled trillium lattices of K2Ni2(SO4)3}

\author{Ivica {\v Z}ivkovi{\' c}*}
\email{ivica.zivkovic@epfl.ch}
\affiliation{Laboratory for Quantum Magnetism, Institute of Physics, \'Ecole Polytechnique F\'ed\'erale de Lausanne, CH-1015 Lausanne, Switzerland}

\author{Virgile Favre}
\affiliation{Laboratory for Quantum Magnetism, Institute of Physics, \'Ecole Polytechnique F\'ed\'erale de Lausanne, CH-1015 Lausanne, Switzerland}

\author{Catalina Salazar Mejia}
\affiliation{Hochfeld-Magnetlabor Dresden (HLD-EMFL) and W\"urzburg-Dresden Cluster of Excellence ct.qmat, Helmholtz-Zentrum Dresden-Rossendorf, 01328 Dresden, Germany}

\author{Harald O. Jeschke}
\affiliation{Research Institute for Interdisciplinary Science, Okayama University, Okayama 700-8530, Japan}

\author{Arnaud Magrez}
\affiliation{Crystal Growth Facility, \'Ecole Polytechnique Fédérale de Lausanne, Lausanne, Switzerland.}

\author{Bhupen Dabholkar}
\affiliation{Department of Physics and Quantum Centers in Diamond and Emerging Materials (QuCenDiEM) group, Indian Institute of Technology Madras, Chennai 600036, India}

\author{Vincent Noculak}
\affiliation{Dahlem Center for Complex Quantum Systems and Fachbereich Physik, Freie Universit\"at Berlin, 14195 Berlin, Germany}
\affiliation{Helmholtz-Zentrum für Materialien und Energie, Hahn-Meitner-Platz 1, 14109 Berlin, Germany}

\author{Rafael S. Freitas}
\affiliation{7 Instituto de F{\' i}sica, Universidade de S{\~ a}o Paulo, 05508-090, S{\~ a}o Paulo, Brazil}

\author{Minki Jeong}
\affiliation{School of Physics and Astronomy, University of Birmingham, Edgbaston, Birmingham, B15 2TT UK}

\author{Nagabhushan G. Hegde}
\affiliation{Laboratory for Quantum Magnetism, Institute of Physics, \'Ecole Polytechnique F\'ed\'erale de Lausanne, CH-1015 Lausanne, Switzerland}

\author{Luc Testa}
\affiliation{Laboratory for Quantum Magnetism, Institute of Physics, \'Ecole Polytechnique F\'ed\'erale de Lausanne, CH-1015 Lausanne, Switzerland}

\author{Peter Babkevich}
\affiliation{Laboratory for Quantum Magnetism, Institute of Physics, \'Ecole Polytechnique F\'ed\'erale de Lausanne, CH-1015 Lausanne, Switzerland}

\author{Yixi Su}
\affiliation{J\"ulich Centre for Neutron Science (JCNS) at Heinz Maier-Leibnitz Zentrum (MLZ), Forschungszentrum J\"ulich, Lichtenbergstrasse 1, D-85747 Garching, Germany}

\author{Pascal Manuel}
\affiliation{ISIS Pulsed Neutron and Muon Source, STFC Rutherford Appleton Laboratory, Harwell Science and Innovation Campus, Didcot, Oxfordshire, OX11 0QX UK}

\author{Hubertus Luetkens}
\affiliation{Laboratory for Muon Spin Spectroscopy, Paul Scherrer Institute, CH-5232 Villigen, Switzerland}

\author{Christopher Baines}
\affiliation{Laboratory for Muon Spin Spectroscopy, Paul Scherrer Institute, CH-5232 Villigen, Switzerland}

\author{Peter J. Baker}
\affiliation{ISIS Pulsed Neutron and Muon Source, STFC Rutherford Appleton Laboratory, Harwell Science and Innovation Campus, Didcot, Oxfordshire, OX11 0QX UK}

\author{Jochen Wosnitza}
\affiliation{Hochfeld-Magnetlabor Dresden (HLD-EMFL) and W\"urzburg-Dresden Cluster of Excellence ct.qmat, Helmholtz-Zentrum Dresden-Rossendorf, 01328 Dresden, Germany}
\affiliation{Institut f\"ur Festk\"orper- und Materialphysik, TU Dresden, 01062 Dresden Germany}

\author{Oksana Zaharko}
\affiliation{Laboratory for Neutron Scattering and Imaging, Paul Scherrer Institut, CH-5253 Villigen, Switzerland}

\author{Yasir Iqbal}
\affiliation{Department of Physics and Quantum Centers in Diamond and Emerging Materials (QuCenDiEM) group, Indian Institute of Technology Madras, Chennai 600036, India}

\author{Johannes Reuther}
\affiliation{Dahlem Center for Complex Quantum Systems and Fachbereich Physik, Freie Universit\"at Berlin, 14195 Berlin, Germany}
\affiliation{Helmholtz-Zentrum für Materialien und Energie, Hahn-Meitner-Platz 1, 14109 Berlin, Germany}

\author{Henrik M. R\o{}nnow}
\affiliation{Laboratory for Quantum Magnetism, Institute of Physics, \'Ecole Polytechnique F\'ed\'erale de Lausanne, CH-1015 Lausanne, Switzerland}

\date{\today}

\begin{abstract}
{Quantum spin liquids are exotic states of matter which form when strongly frustrated magnetic interactions induce a highly entangled quantum paramagnet far below the energy scale of the magnetic interactions. Three-dimensional cases are especially challenging due to the significant reduction of the influence of quantum fluctuations. Here, we report the magnetic characterization of {\kni} forming a three dimensional network of Ni$^{2+}$ spins. Using density functional theory calculations we show that this network consists of two interconnected spin-1 trillium lattices. In the absence of a magnetic field, magnetization, specific heat, neutron scattering and muon spin relaxation experiments demonstrate a highly correlated and dynamic state, coexisting with a peculiar, very small static component exhibiting a strongly renormalized moment. A magnetic field $B \gtrsim 4$ T diminishes the ordered component and drives the system in a pure quantum spin liquid state. This shows that a system of interconnected $S=1$ trillium lattices exhibit a significantly elevated level of geometrical frustration.}
    
\end{abstract}

\maketitle

Strongly correlated systems are at the forefront of condensed matter research, exhibiting exotic phases and nourishing novel theoretical concepts. In magnetism, one of the most sought-after strongly correlated phase is a quantum spin liquid (QSL), a state in which spins avoid long-range order (LRO) and are considered entangled on all spatial scales~\cite{Savary-2016,Zhou-2017,Broholm-2020}. To realize a QSL, geometrical frustration and reduced dimensionality of the magnetic subsystem have been considered vital. 1D Heisenberg chains exhibit QSL behavior even without frustration~\cite{Lake2013,Mourigal2013} while 3D cases are rare due to the significant reduction of quantum fluctuations. Nevertheless, it has been found that 3D lattices like pyrochlore~\cite{Wen2017,Gao2019,Plumb2019} and hyper-hyperkagome~\cite{Okamoto2007,Chillal2020} support QSL behavior.

In this Letter we provide extensive experimental and computational evidence that {\kni} exhibits QSL behavior, based on a novel arrangement of spins forming two interconnected trillium lattices. Previous work on compounds featuring a {\it single} trillium lattice was mainly driven by a pressure-induced quantum phase transition (QPT) discovered in the itinerant helimagnet MnSi~\cite{Pfleiderer1997} and evidence of non-Fermi liquid behavior above a critical pressure~\cite{Pfleiderer2001}. Later theoretical works~\cite{Hopkinson2006,Isakov2008} showed some degree of geometrical frustration in the trillium lattice, nevertheless insufficient to prevent the onset of LRO. From that perspective, {\kni} and other members of the langbeinite family K$_2${\it M}$_2$(SO$_4$)$_3$ ({\it M} = Fe, Co, Mn, Cr) offer an arena for testing future theoretical developments on interconnected trillium lattices. Previous investigations of those compounds displayed ferroelectricity and structural transitions but their magnetic properties remain {\it terra incognita}.

\begin{figure*}
\centering
\includegraphics[width=0.9\columnwidth]{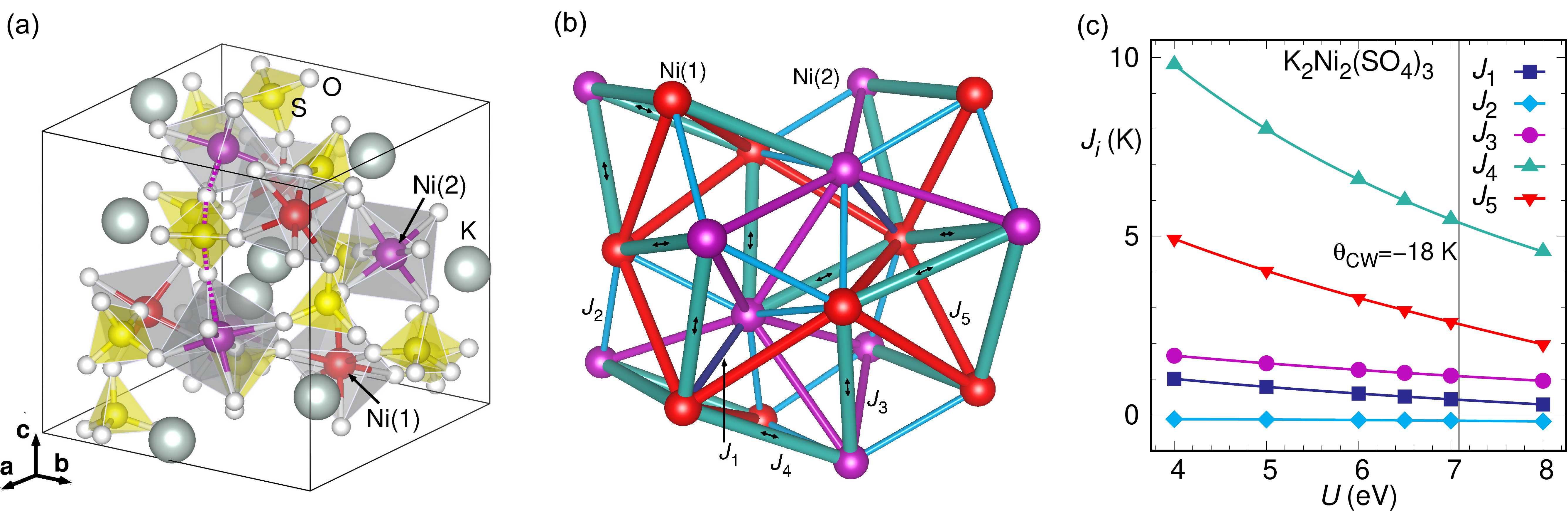}
\caption{(a) Unit cell of {\kni}. A Ni--O--S--O--Ni {\it super-super}-exchange path contributing to a trillium coupling is marked by dashed lines. (b) Exchange network between nickel sites. A ten site loop formed by the strongest exchange $J_4$ is marked by arrows. (c) Exchange couplings determined by DFT energy mapping. The vertical line indicates the $U$ value where the calculated Curie-Weiss temperature matches the experimental value.}
\label{fig:structure}
\end{figure*}

{\kni} crystallizes in a cubic unit cell ($P2_13$) with $a = 9.81866(12)$ \AA~determined from single-crystal diffraction at 100\,K~\cite{SM}. It consists of a network of trigonally-distorted NiO$_6$ octahedra, coupled through SO$_4$ groups [Fig.~\ref{fig:structure}(a)], with a Ni--O--S--O--Ni {\it super-super}-exchange mechanism mediating magnetic interactions between $S = 1$ spins. There are two crystallographic Ni sites, distinguished by their Ni--O distances~\cite{SM}, each site forming a single trillium lattice.

Mapping the GGA+$U$ total energies~\cite{SM} onto a Heisenberg Hamiltonian for {\kni} $\hat{\cal H}=\sum_{i<j} J_{ij} {\bf{\hat S}}_i\cdot {\bf{\hat S}}_j$ as shown in Fig.~\ref{fig:structure}(c) yields the five non-zero exchange couplings that are listed in Table~\ref{table:exchange_paths} and shown in Fig.~\ref{fig:structure}(b), visualizing the exchange network. The couplings within each TL are given by antiferromagnetic (AFM) $J_3$ and $J_5$, respectively. On the other hand the strongest coupling is found to be AFM $J_4$ that inter-connects the two lattices. Interestingly, if $J_4$ was the only coupling in the system it would support a N\'eel-type LRO. Thus, our calculation shows that the physics of {\kni} is determined by an interplay between $J_4$ induced ordering tendencies and $J_3$ and $J_5$ driven frustration.

\begin{table}
\begin{tabular}{lccc}
\hline\hline
Label & Type & Distance (\AA) & Exchange (K) \\
\hline
$J_1$ & Ni(1)--Ni(2) & 4.42877 & 0.42(1)\\
$J_2$ & Ni(1)--Ni(2) & 4.90057 & $-0.16(1)$ \\
$J_3$ & Ni(2)--Ni(2) & 6.08379 & 1.09(1) \\
$J_4$ & Ni(1)--Ni(2) & 6.12050 & 5.38(1) \\
$J_5$ & Ni(1)--Ni(1) & 6.12695 & 2.54(1) \\
\hline\hline
\end{tabular}
\caption{{\kni} exchange energies obtained by DFT energy mapping, with paths identified by Ni--Ni distance. }
\label{table:exchange_paths}
\end{table}

\begin{figure}
    \centering
    \includegraphics[width=0.7\columnwidth]{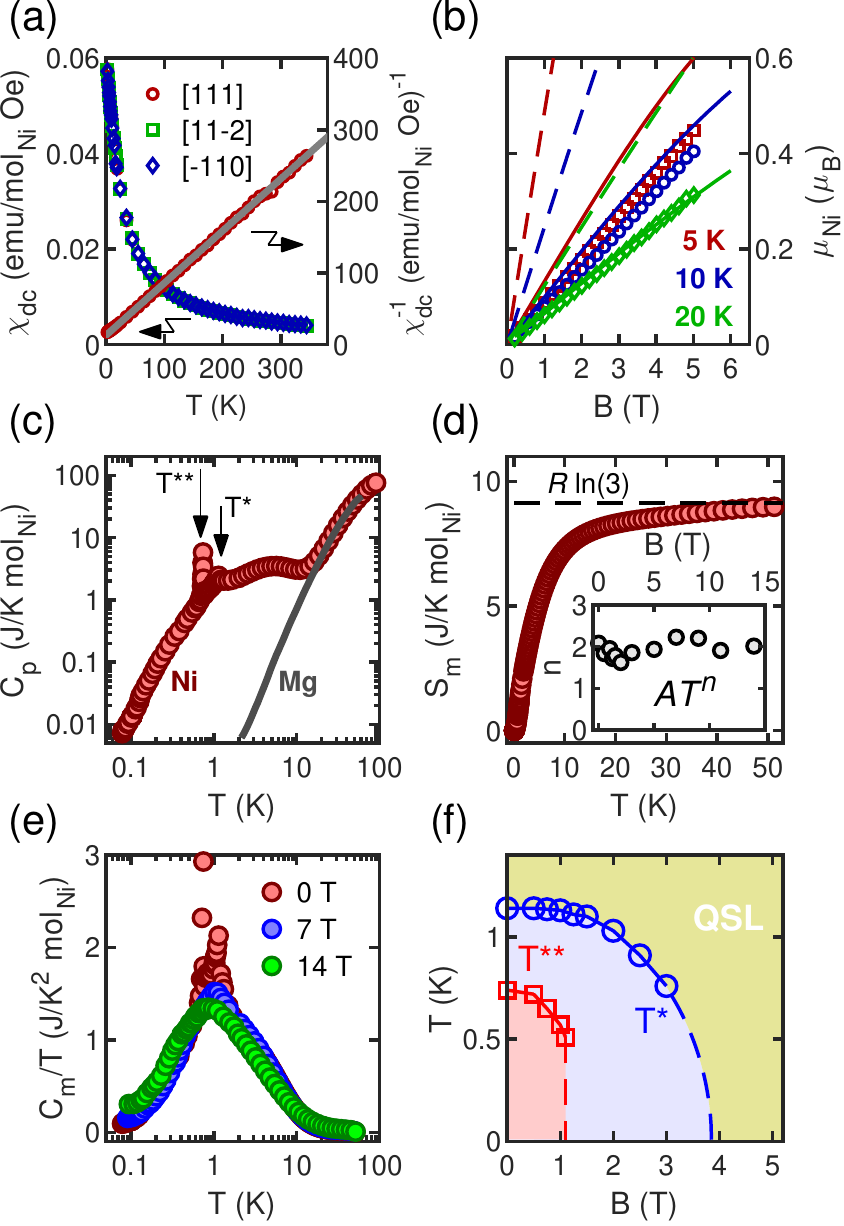}
    \caption{(a) $T$ dependence of $\chi_{\rm dc}$ along the three orthogonal directions (left axis) for $B = 0.1$ T, and the inverse susceptibility $1/\chi_{\rm dc}$ (right axis) with $B || [111]$. The solid grey line represents the Curie-Weiss law. (b) Magnetic field dependence of magnetization, together with the Brillouin function 
    for $S = 1$ and $g = 2.34$ (dashed lines) and Monte Carlo simulations (solid lines). (c) Zero field temperature dependence of the total specific heat of {\kni}, together with the total specific heat of {\kmg} representing the phonon contribution. (d) $T$ dependence of magnetic entropy. An inset shows the magnetic field dependence of the exponent $n$ in the power law $C_p = AT^n$. (e) $T$ dependence of the magnetic specific heat for several magnetic field values. (f) $T$ -- $B$ phase diagram of {\kni}.}
    \label{fig:thermodynamics}
\end{figure}

Figure~\ref{fig:thermodynamics}(a) displays the temperature dependence of dc magnetic susceptibility $\chi_{\rm dc} (T)$ and its inverse in a wide temperature range. The monotonic increase of $\chi_{\rm dc} (T)$ with decreasing $T$, without any noticeable features, suggests the absence of LRO down to 2\,K. The linear behavior of $1/\chi_{\rm dc} (T)$ above 50\,K allows us to use the Curie-Weiss law $\chi (T) = C/(T - \Theta_{\rm CW})$, which gives $C = 1.37(2)$ emu\,K/mol and $\Theta_{\rm CW} = -18(1)$\,K. The value of $C$ corresponds to $S = 1$ with a slightly enhanced $g$-factor of $g = 2.34$. $\Theta_{\rm CW} < 0$ indicates predominant AFM interactions, in accordance with density functional theory (DFT) calculations. Additionally, measurements along three orthogonal directions practically overlap, indicating no significant anisotropy.

Below 50\,K, $1/\chi_{\rm dc} (T)$ starts to deviate from the Curie-Weiss law, following the build up of correlations between magnetic moments. To emphasize this behavior, magnetization curves obtained at several temperatures are plotted in Fig.~\ref{fig:thermodynamics}(b), together with the curves of the Brillouin function, which describe an assembly of non-interacting $S = 1$ spins, at corresponding temperatures (dashed lines). To approximate the magnetization behavior, classical Monte Carlo calculations employing the DFT Hamiltonian have been performed. The resulting curves (full lines) are closer to the experimental ones but it is apparent that with decreasing $T$ the deviation from the classical prediction becomes more pronounced, suggesting a sizeable influence of quantum fluctuations on this 3D lattice.

Further evidence of strongly correlated spins can be obtained from specific heat measurements. Fig.~\ref{fig:thermodynamics}(c) shows the $T$ dependence of the total specific heat of a single crystal of {\kni}, together with a non-magnetic analog {\kmg}. At temperatures above 20\,K the two compounds show a very similar behavior, indicating a dominant phonon contribution. Below 20\,K, {\kni} exhibits a significant deviation, with a broad maximum around 5\,K and two features occurring at $T^* = 1.14$\,K and $T^{**} = 0.74$\,K. Below $T^{**}$, the heat capacity behaves according to a power-law $C_p \sim T^n$, with $n \approx 2$. This value of the exponent differs appreciably from $n=3$ for classical AFM and has been observed in several frustrated magnetic systems~\cite{Ramirez2000,Nakatsuji2005,Silverstein2014,Plumb2019}.

To extract the magnetic specific heat $C_m$, the phonon contribution using the data obtained on {\kmg} has been subtracted. On the high $T$ side the subtraction works up to 50\,K where {\kmg} shows a kink~\cite{SM}, associated to previously observed lattice related features in the heat capacity~\cite{Boerio1990}. On the low $T$ side a polynomial $BT^3 + CT^5$ has been used~\cite{SM}. The error of the total entropy $S = \int (C_m/T)dT$ due to background subtraction is estimated to be a few percent. As can be seen in Fig.~\ref{fig:thermodynamics}(d), at 50\,K more than 98\% of the expected entropy for $S = 1$ system is recovered, with more than 90\% being released up to 20\,K. The saturation towards the $R\ln{(2S+1)}$ value for $S = 1$ indicates that no residual entropy is present at $T = 0$ and that {\kni} exhibits a non-degenerate ground state.

Application of magnetic field along the [111] direction induces little change in the overall behavior of the heat capacity of {\kni}. A small but noticeable redistribution occurs for fields above $B = 7$ T [ Fig.~\ref{fig:thermodynamics}(e)] but even with fields up to 14 T the overall shape of the curve remains unchanged. The power law $C_p \sim T^n$ observed at low $T$ for $B = 0$ is maintained for $B > 0$ without a visible crossover towards the gapped polarized state, as seen, for example, in YbMgGaO$_4$~\cite{Xu2016}. The value of the extracted exponent remains field independent up to 14\,T [inset of Fig.~\ref{fig:thermodynamics}(d)].

The order of transitions at $T^{*}$ and $T^{**}$ is revealed through their overall shape. The feature at $T^{*}$ resembles a typical, asymmetric $\lambda$-shape, characteristic of second-order phase transitions. On the other hand, at $T^{**}$ a narrow, symmetrical peak is found, often seen in first-order phase transitions. Although the entropy released at $T^*$ amounts to only 1\% of the total $R{\rm ln}3$~\cite{SM}, the sample purity determined by single-crystal x-ray diffraction~\cite{SM} rules out any impurity-related scenario. Additionally, a comparison with specific-heat measurements on a powder sample reveals that $T^{*}$ is significantly diminished while $T^{**}$ is completely absent~\cite{SM}. With a tentative assignment of $T^{**}$ as a first-order phase transition, its presence in a single-crystal experiment suggests that it is intrinsically related to the low-temperature magnetic phase of {\kni}.

The magnetic field dependence of $T^{*}$ and $T^{**}$ is presented in Fig.~\ref{fig:thermodynamics}(f). $T^{**}$ is quickly diminished in amplitude and for $B > 1$T it disappears completely. $T^{*}$ is practically unchanged up to $B = 1$ T with a subsequent decrease and a reduction of the size of the anomaly~\cite{SM}. By assuming a quadratic $B$-dependence of the second-order phase transition the value of the critical magnetic field $B_c \lesssim 4$ T can been estimated, above which a completely dynamic and fluctuating state exists down to the lowest $T$.

\begin{figure}
    \centering
    \includegraphics[width=0.8\columnwidth]{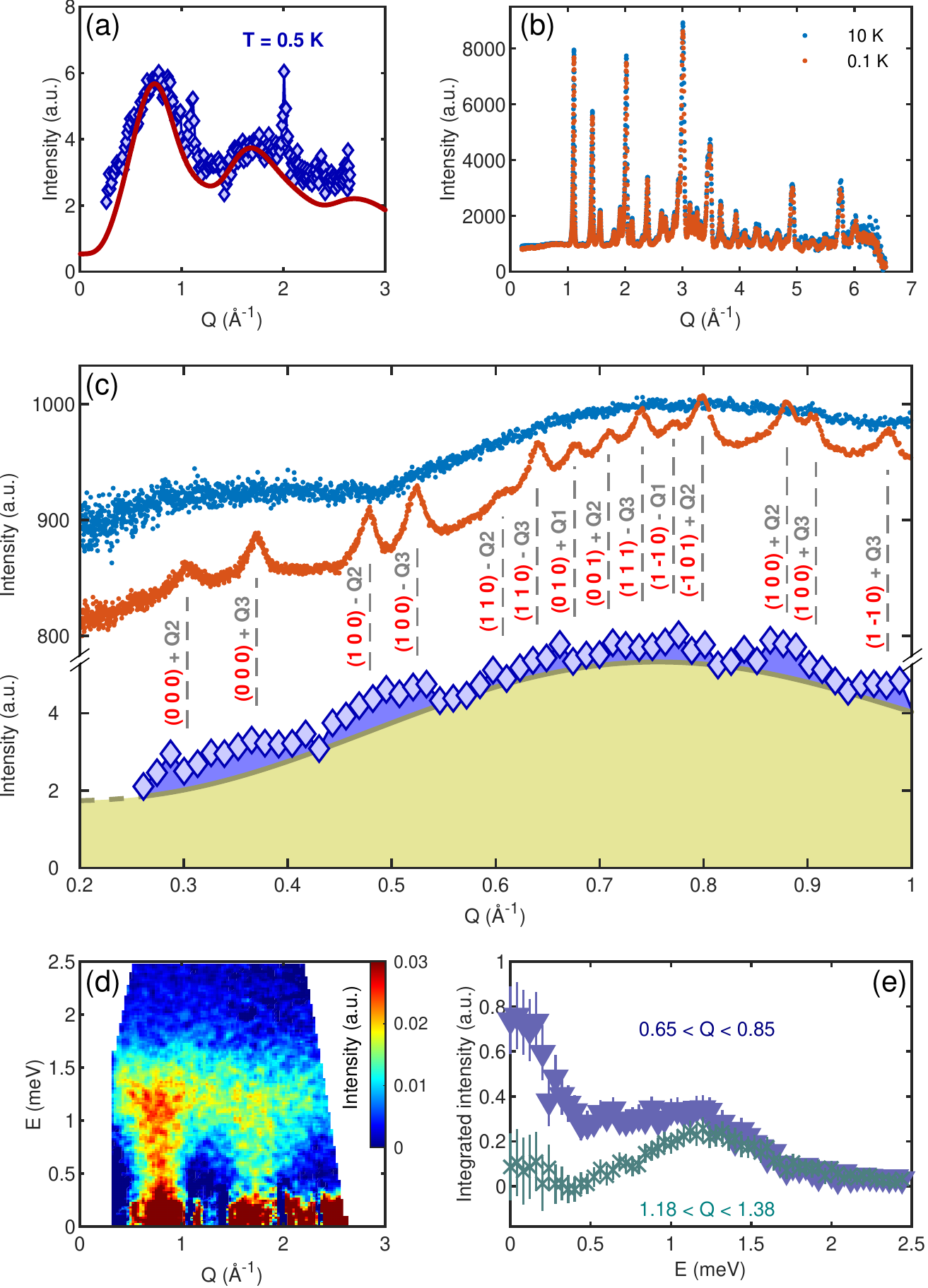}
    \caption{(a) Pure magnetic scattering pattern derived from the spin-polarized neutron diffraction data at 0.5\,K (blue diamonds) and from PFFRG (solid line), (b) Neutron powder diffraction at 0.1\,K and 10\,K, (c) Combined data from polarized neutrons (blue diamonds from panel (a)) and powder diffraction (blue (10\,K) and red points (0.1\,K) from panel (b)). Blue shading indicates an upper limit of the contribution from the ordered component, yellow shading represents the contribution from the fluctuating component, (d) TOF data at 0.5\,K, (e) $E$-dependence of the integrated intensity for two $Q$-ranges.}
    \label{fig:neutrons}
\end{figure}

To shed more light on the peculiar magnetic properties of {\kni}, a series of neutron scattering experiments have been performed. Fig.~\ref{fig:neutrons}(a) shows the results of polarized neutron scattering, in which a $Q$-dependence of the scattering intensity at 0.5\,K is presented. It exhibits a broad maximum centered at $Q_{\rm max} \approx 0.75$ \AA$^{-1}$ followed by an attenuating oscillatory dependence. Such a broad, liquid-like structure factor is typical for systems with strong quantum fluctuations. This conclusion is further supported by the fact that the diffuse scattering pattern in Fig.~\ref{fig:neutrons}(a) is well reproduced by pseudofermion functional renormalization group (PFFRG) simulations of the DFT Hamiltonian. Remarkably, despite the general difficulties in simulating a strongly fluctuating 3D spin system with complex frustrated interactions as realized in K$_2$Ni$_2$(SO$_4$)$_3$, not only the positions of the extrema are well reproduced but also the global amplitude variations. Additionally, the oscillatory behavior is seen to persist at least up to 17 K~\cite{SM}, clearly indicating its connection to strong correlations developing below 20 K.

To investigate the system's static component, a neutron powder diffraction experiment has been performed well above and well below $T^{*}$. From a wide $Q$ diffraction pattern shown in Fig.~\ref{fig:neutrons}(b) it is found that for $Q > 1$ \AA$^{-1}$ all peaks are present at both temperatures, indicating their lattice origin. On the other hand, a series of very weak magnetic peaks can be found at $T = 0.1$\,K for $Q < 1$ \AA$^{-1}$ as seen in the upper half of Fig.~\ref{fig:neutrons}(c). They can all be assigned to satellites of the main nuclear Bragg peaks $(h,k,l)$ in the form $q_{\rm magnetic} = (h,k,l) \pm Q_i$, where $Q_1 = (\frac{1}{3},0,0)$, $Q_2 = (\frac{1}{3},\frac{1}{3},0)$ and $Q_3 = (\frac{1}{3},\frac{1}{3},\frac{1}{3})$. The existence of three propagation vectors indicates that even LRO is heavily influenced by frustration, leaving several possible structures with similar ground state energies.

Due to the complexity of the scattering pattern, including several propagation vectors, tripling of the magnetic unit cell and very weak amplitudes, it is not possible to completely determine the magnetic structure nor to extract the value of the ordered moment. Nevertheless, utilizing a purely magnetic scattering pattern from polarized neutrons [Fig.~\ref{fig:neutrons}(a)] one can estimate an upper limit for the static component. To this end, we envisage that the total intensity $S(Q)$ is composed of two contributions $S(Q) = S_{\rm static}(Q) + S_{\rm dynamic}(Q)$, with the jagged $S_{\rm static}(Q)$ roughly following the powder diffraction profile and sitting on top of the smooth $S_{\rm dynamic}(Q)$. Although the resultant ratio $S_{\rm static}(Q)/S(Q) \approx 11\%$ cannot be directly related to the value of the ordered moment, it serves as a supporting evidence that the ground state in {\kni} is dominated by spin fluctuations.

In Fig.~\ref{fig:neutrons}(d), we show time-of-flight (TOF) data obtained as a direct subtraction of the background intensity obtained at 80\,K from a measured intensity at 0.5\,K. Streaks of intensity can be observed at the same positions as maxima in $S(Q)$ found with polarized neutrons. The upper limit of spin excitations is found to be around 1.8 meV which agrees well with the temperature at which specific heat starts to significantly deviate from a purely phononic behavior. In Fig.~\ref{fig:neutrons}(e), a narrow Q-integrated energy dependence of intensity is shown, indicating a continuum of excitations down to the elastic line. Due to the existence of the ordered component, it is not straight-forward to assign this continuum to the QSL state. On the other hand, the dominance of the dynamic component, revealed by specific heat data and polarized neutron scattering, renders this conclusion very plausible, which would then support the hypothesis of a gapless nature for the QSL.

\begin{figure}
    \centering
    \includegraphics[width=0.9\columnwidth]{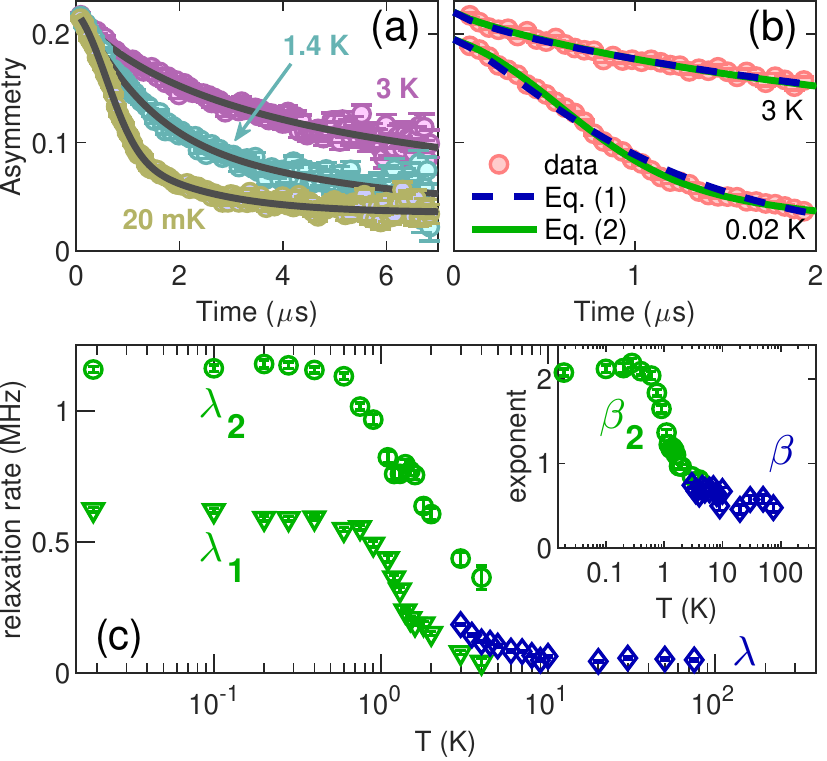}
    \caption{(a) ZF-$\mu$SR spectra at selected temperatures. The solid lines represent fitted curves following the model in Eq.~\eqref{eq:2contr}. (b) A comparison of the applicability of two models at 0.02 K (shifted down for clarity) and 3 K. (c) Temperature dependence of the $\mu$SR rates extracted from two models. Circles and triangles represent relaxation rates from the model in Eq.~\eqref{eq:2contr}, diamonds from the model in Eq.~\eqref{eq:stretch}. The inset displays the temperature dependence of the exponent $\beta_2$ from Eq.~\eqref{eq:2contr} and $\beta$ from Eq.~\eqref{eq:stretch}.}
    \label{fig:muSR}
\end{figure}

To probe further the peculiar coexistence of static and dynamic properties revealed in {\kni}, muon spin relaxation ($\mu$SR) experiments have been performed. As shown in Fig.~\ref{fig:muSR}(a), no obvious wiggles are observed down to lowest $T$. On a phenomenological level the relaxation is often described by a stretched-exponential function
\begin{equation}
    A(t) = A_0 e^{-(\lambda t)^{\beta}} + A_{\rm BG},
\label{eq:stretch}
\end{equation}
where $A_0$ is the initial asymmetry, $A_{\rm BG}$ a constant background, $\lambda$ is the relaxation rate and $\beta$ is the stretching exponent that in an ideal case of $\beta = 1$ leads to a simple exponential relaxation. $\beta < 1$ has usually been associated with either a distribution of relaxation times, multiple muon stopping sites, or with intrinsic disorder in the magnetic system. As is evident from Fig.~\ref{fig:muSR}(b), at low $T$ the observed time dependence of the asymmetry cannot be satisfactorily described by a single contribution. Thus, we have extended Eq.~\eqref{eq:stretch} with an additional term
\begin{equation}
    A(t) = A_0 (f e^{-(\lambda_1 t)^{\beta_1}} + (1-f) e^{-(\lambda_2 t)^{\beta_2}}) + A_{\rm BG}.
\label{eq:2contr}
\end{equation}
and fixed $f = 0.5$ and $\beta_1 = 1$ to avoid over-parametrization. We find that it is necessary to use Eq.~\eqref{eq:2contr} up to 3 K while for $T > 3$ K Eq.~\eqref{eq:stretch} is sufficient (for the discussion of the overlapping region see~\cite{SM}). In Fig.~\ref{fig:muSR}(c), we present the temperature evolution of relaxation rates and exponents (see inset) extracted using Eq.~\eqref{eq:2contr} (green symbols) and Eq.~\eqref{eq:stretch} (blue symbols).

Below $T \sim 1$\,K, the extracted parameters attain a constant value, a feature often associated with a highly dynamic nature of QSLs~\cite{Mendels2007,Balz16,Fujihala2020}. We point out that the value of the exponent $\beta \simeq 2$ is indicative of a specific type of a correlated spin system based on spin-singlets~\cite{Uemura1994}. Within this scenario, the Gaussian shape of the relaxation profile develops from a sporadic appearance of unpaired spins. The time interval of their existence is much shorter than a life-time of a muon, so for the majority of time muons experience very small fields related to the short-lived but very distant unpaired spins. Such a scenario is in accordance with a practically field-independent magnetic specific heat seen in Fig.~\ref{fig:thermodynamics}(e)~\cite{Ramirez2000}. Within this framework the strong relaxation at low temperatures described by $\lambda_1$ can be associated with a partial but homogeneous order while the remaining dynamics is due to the sporadic unpaired-spin appearances. The absence of oscillations can then be associated with a spread of local fields originating from complex magnetic structures given by propagation vectors $Q_1$, $Q_2$ and $Q_3$. Additionally, the coherent regions giving rise to magnetic peaks in neutron diffraction are probed on much shorter time scales ($\sim 10^{-14}$ s), allowing for local fluctuations between different magnetic structures on the time scale of muons.

We find two possible scenarios that could encompass a small value of the static component existing alongside the dominant, fluctuating component. The first scenario assumes the existence of a quantum critical point (QCP) between an ordered phase and a quantum-fluctuation-dominant phase, with {\kni} being on the ordered side of QCP but ``accidentally'' close to it. In this case, the ordered moment $m_s$ is strongly renormalized due to the prevalence of quantum fluctuations close to a QCP, as has been demonstrated in TlCuCl$_3$ where a pressure-controlled QPT between a LRO AFM state and a non-magnetic dimer phase is arbitrarily decreased ($m_{s} \sim \sqrt{p - p_c}$) close to a QCP~\cite{Ruegg2008}. In this context, a possible control parameter could be the ratio of intra- ($J_3$, $J_5$) and inter-trillium lattice couplings ($J_1$, $J_2$, $J_4$). Given that Ni(1) and Ni(2) sites form a bipartite lattice, the limit of dominant $J_1$, $J_2$, $J_4$ results in a semiclassical AFM phase. With $J_3$, $J_5$ dominant, the system is in the limit of two weakly coupled trillium lattices. As demonstrated theoretically for a single trillium lattice, it is expected to form a variant of the 120$\degree$ order~\cite{Hopkinson2006,Isakov2008}. The case of two interconnected trillium lattices represents a novel research direction with many members of the langbeinite family providing ample opportunity for comparison with theory.

The second scenario dismisses the ``fortuitous'' constellation of parameters describing {\kni} and considers it positioned well within the QSL phase. Due to the presence of antisymmetric exchange coupling (the Dzyaloshinskii-Moriya interaction (DMI)) allowed by the non-centrosymmetric space group, the ground state gets ``dressed'' with a small ordered component due to the admixing of higher lying states, similar to the admixture of triplet wave-functions into the ground state singlet of an AFM dimer. An exciting consequence of this scenario arises from topological aspects imposed on the QSL state. Magnetic structures forming in non-centrosymmetric space groups are shown to support skyrmions, topologically protected spin textures ~\cite{Muhlbauer2009,Seki2012}. Fractional wave-numbers $Q_1$, $Q_2$ and $Q_3$ revealed in the diffraction experiment do indicate a potential role of DMI in the formation of LRO.

In either case, the observed coexistence between fluctuating spins and a small static component which vanishes in a magnetic field could be linked to already developed concepts like field-induced spin liquids in Kitaev-type honeycomb models featuring non-Abelian fractional quasiparticles~\cite{Kitaev2006}. The ability to tune its behavior across QCP with magnetic field into a pristine QSL state is an exciting opportunity which should stimulate further experimental and theoretical studies.

{\it Acknowledgments.} We thank Bi Wen Hua for his help with x-ray diffraction experiments. I. {\v Z}. acknowledges financial support by the Swiss National Science Foundation (SNSF) project No.~200021-169699 and 206021-189644. H. M. R. acknowledges financial support by SNFS projects No.~200020-188648 and 206021-189644. Part of this work was supported by the Deutsche Forschungsgemeinschaft (DFG) through the W\"urzburg-Dresden Cluster of Excellence on Complexity and Topology in Quantum Matter--$ct:qmat$ (EXC 2147, Project No. 390858490), as well as by the HLD at HZDR, a member of the European Magnetic Field Laboratory (EMFL). R.S.F. acknowledges financial support by FAPESP (Grant No. 2015/16191-5) and CNPq (Grant No. 429511/2018-3). J.R. acknowledges financial support by the German Research Foundation within the CRC183 (project A04). Y.I. acknowledges financial support by SERB, Department of Science and Technology (DST), India through grants SRG (No.~SRG/2019/000056), MATRICS (No.~MTR/2019/001042), and Indo-French Center for the Promotion of Advanced Research CEFIPRA (No. 64T3-1). This research was supported in part by the National Science Foundation under Grant No.~NSF~PHY-1748958, the ICTP through the Simons Associateship scheme, IIT Madras through the IoE program for establishing the QuCenDiEM group (Project No. SB20210813PHMHRD002720), ICTS, Bengaluru, India during a visit for participating in the program “Novel phases of quantum matter” (Code: ICTS/topmatter2019/12). Y.I. acknowledges the use of the computing resources at HPCE, IIT Madras. The $\mu$SR experiments were performed at MUSR beamline, ISIS (1710223) and LTF and GPS beamlines, PSI (2017119 and 2017119). The neutron diffraction experiment was performed at WISH beamline, ISIS (2010010). The spin-polarized neutron diffraction and TOF experiments were performed at DNS beamline, MLZ (13656).

\bibliography{ref}

\begin{thebibliography}{38}%
\makeatletter
\providecommand \@ifxundefined [1]{%
 \@ifx{#1\undefined}
}%
\providecommand \@ifnum [1]{%
 \ifnum #1\expandafter \@firstoftwo
 \else \expandafter \@secondoftwo
 \fi
}%
\providecommand \@ifx [1]{%
 \ifx #1\expandafter \@firstoftwo
 \else \expandafter \@secondoftwo
 \fi
}%
\providecommand \natexlab [1]{#1}%
\providecommand \enquote  [1]{``#1''}%
\providecommand \bibnamefont  [1]{#1}%
\providecommand \bibfnamefont [1]{#1}%
\providecommand \citenamefont [1]{#1}%
\providecommand \href@noop [0]{\@secondoftwo}%
\providecommand \href [0]{\begingroup \@sanitize@url \@href}%
\providecommand \@href[1]{\@@startlink{#1}\@@href}%
\providecommand \@@href[1]{\endgroup#1\@@endlink}%
\providecommand \@sanitize@url [0]{\catcode `\\12\catcode `\$12\catcode
  `\&12\catcode `\#12\catcode `\^12\catcode `\_12\catcode `\%12\relax}%
\providecommand \@@startlink[1]{}%
\providecommand \@@endlink[0]{}%
\providecommand \url  [0]{\begingroup\@sanitize@url \@url }%
\providecommand \@url [1]{\endgroup\@href {#1}{\urlprefix }}%
\providecommand \urlprefix  [0]{URL }%
\providecommand \Eprint [0]{\href }%
\providecommand \doibase [0]{https://doi.org/}%
\providecommand \selectlanguage [0]{\@gobble}%
\providecommand \bibinfo  [0]{\@secondoftwo}%
\providecommand \bibfield  [0]{\@secondoftwo}%
\providecommand \translation [1]{[#1]}%
\providecommand \BibitemOpen [0]{}%
\providecommand \bibitemStop [0]{}%
\providecommand \bibitemNoStop [0]{.\EOS\space}%
\providecommand \EOS [0]{\spacefactor3000\relax}%
\providecommand \BibitemShut  [1]{\csname bibitem#1\endcsname}%
\let\auto@bib@innerbib\@empty
\bibitem [{\citenamefont {Savary}\ and\ \citenamefont
  {Balents}(2016)}]{Savary-2016}%
  \BibitemOpen
  \bibfield  {author} {\bibinfo {author} {\bibfnamefont {L.}~\bibnamefont
  {Savary}}\ and\ \bibinfo {author} {\bibfnamefont {L.}~\bibnamefont
  {Balents}},\ }\bibfield  {title} {\bibinfo {title} {{Quantum spin liquids: a
  review}},\ }\href {https://doi.org/10.1088/0034-4885/80/1/016502} {\bibfield
  {journal} {\bibinfo  {journal} {Rep. Prog. Phys.}\ }\textbf {\bibinfo
  {volume} {80}},\ \bibinfo {pages} {016502} (\bibinfo {year}
  {2016})}\BibitemShut {NoStop}%
\bibitem [{\citenamefont {Zhou}\ \emph {et~al.}(2017)\citenamefont {Zhou},
  \citenamefont {Kanoda},\ and\ \citenamefont {Ng}}]{Zhou-2017}%
  \BibitemOpen
  \bibfield  {author} {\bibinfo {author} {\bibfnamefont {Y.}~\bibnamefont
  {Zhou}}, \bibinfo {author} {\bibfnamefont {K.}~\bibnamefont {Kanoda}},\ and\
  \bibinfo {author} {\bibfnamefont {T.-K.}\ \bibnamefont {Ng}},\ }\bibfield
  {title} {\bibinfo {title} {{Quantum spin liquid states}},\ }\href
  {https://doi.org/10.1103/RevModPhys.89.025003} {\bibfield  {journal}
  {\bibinfo  {journal} {Rev. Mod. Phys.}\ }\textbf {\bibinfo {volume} {89}},\
  \bibinfo {pages} {025003} (\bibinfo {year} {2017})}\BibitemShut {NoStop}%
\bibitem [{\citenamefont {Broholm}\ \emph {et~al.}(2020)\citenamefont
  {Broholm}, \citenamefont {Cava}, \citenamefont {Kivelson}, \citenamefont
  {Nocera}, \citenamefont {Norman},\ and\ \citenamefont
  {Senthil}}]{Broholm-2020}%
  \BibitemOpen
  \bibfield  {author} {\bibinfo {author} {\bibfnamefont {C.}~\bibnamefont
  {Broholm}}, \bibinfo {author} {\bibfnamefont {R.~J.}\ \bibnamefont {Cava}},
  \bibinfo {author} {\bibfnamefont {S.~A.}\ \bibnamefont {Kivelson}}, \bibinfo
  {author} {\bibfnamefont {D.~G.}\ \bibnamefont {Nocera}}, \bibinfo {author}
  {\bibfnamefont {M.~R.}\ \bibnamefont {Norman}},\ and\ \bibinfo {author}
  {\bibfnamefont {T.}~\bibnamefont {Senthil}},\ }\bibfield  {title} {\bibinfo
  {title} {{Quantum spin liquids}},\ }\href
  {https://doi.org/10.1126/science.aay0668} {\bibfield  {journal} {\bibinfo
  {journal} {Science}\ }\textbf {\bibinfo {volume} {367}},\ \bibinfo {pages}
  {6475} (\bibinfo {year} {2020})}\BibitemShut {NoStop}%
\bibitem [{\citenamefont {Lake}\ \emph {et~al.}(2013)\citenamefont {Lake},
  \citenamefont {Tennant}, \citenamefont {Caux}, \citenamefont {Barthel},
  \citenamefont {Schollw\"{o}ck}, \citenamefont {Nagler},\ and\ \citenamefont
  {Frost}}]{Lake2013}%
  \BibitemOpen
  \bibfield  {author} {\bibinfo {author} {\bibfnamefont {B.}~\bibnamefont
  {Lake}}, \bibinfo {author} {\bibfnamefont {D.~A.}\ \bibnamefont {Tennant}},
  \bibinfo {author} {\bibfnamefont {J.~S.}\ \bibnamefont {Caux}}, \bibinfo
  {author} {\bibfnamefont {T.}~\bibnamefont {Barthel}}, \bibinfo {author}
  {\bibfnamefont {U.}~\bibnamefont {Schollw\"{o}ck}}, \bibinfo {author}
  {\bibfnamefont {S.~E.}\ \bibnamefont {Nagler}},\ and\ \bibinfo {author}
  {\bibfnamefont {C.~D.}\ \bibnamefont {Frost}},\ }\bibfield  {title} {\bibinfo
  {title} {{Multispinon Continua at Zero and Finite Temperature in a Near-Ideal
  Heisenberg Chain}},\ }\href {https://doi.org/10.1103/PhysRevLett.111.137205}
  {\bibfield  {journal} {\bibinfo  {journal} {Phys. Rev. Lett.}\ }\textbf
  {\bibinfo {volume} {111}},\ \bibinfo {pages} {137205} (\bibinfo {year}
  {2013})}\BibitemShut {NoStop}%
\bibitem [{\citenamefont {Mourigal}\ \emph {et~al.}(2013)\citenamefont
  {Mourigal}, \citenamefont {Enderle}, \citenamefont {Kl\"{o}pperpieper},
  \citenamefont {Caux}, \citenamefont {Stunault},\ and\ \citenamefont
  {R{\o}nnow}}]{Mourigal2013}%
  \BibitemOpen
  \bibfield  {author} {\bibinfo {author} {\bibfnamefont {M.}~\bibnamefont
  {Mourigal}}, \bibinfo {author} {\bibfnamefont {M.}~\bibnamefont {Enderle}},
  \bibinfo {author} {\bibfnamefont {A.}~\bibnamefont {Kl\"{o}pperpieper}},
  \bibinfo {author} {\bibfnamefont {J.~S.}\ \bibnamefont {Caux}}, \bibinfo
  {author} {\bibfnamefont {A.}~\bibnamefont {Stunault}},\ and\ \bibinfo
  {author} {\bibfnamefont {H.~M.}\ \bibnamefont {R{\o}nnow}},\ }\bibfield
  {title} {\bibinfo {title} {{Fractional spinon excitations in the quantum
  Heisenberg antiferromagnetic chain}},\ }\href
  {https://doi.org/10.1038/nphys2652} {\bibfield  {journal} {\bibinfo
  {journal} {Nat. Phys.}\ }\textbf {\bibinfo {volume} {9}},\ \bibinfo {pages}
  {435} (\bibinfo {year} {2013})}\BibitemShut {NoStop}%
\bibitem [{\citenamefont {Wen}\ \emph {et~al.}(2017)\citenamefont {Wen},
  \citenamefont {Koohpayeh}, \citenamefont {Ross}, \citenamefont {Trump},
  \citenamefont {McQueen}, \citenamefont {Kimura}, \citenamefont {Nakatsuji},
  \citenamefont {Qiu}, \citenamefont {Pajerowski}, \citenamefont {Copley},\
  and\ \citenamefont {Broholm}}]{Wen2017}%
  \BibitemOpen
  \bibfield  {author} {\bibinfo {author} {\bibfnamefont {J.-J.}\ \bibnamefont
  {Wen}}, \bibinfo {author} {\bibfnamefont {S.~M.}\ \bibnamefont {Koohpayeh}},
  \bibinfo {author} {\bibfnamefont {K.~A.}\ \bibnamefont {Ross}}, \bibinfo
  {author} {\bibfnamefont {B.~A.}\ \bibnamefont {Trump}}, \bibinfo {author}
  {\bibfnamefont {T.~M.}\ \bibnamefont {McQueen}}, \bibinfo {author}
  {\bibfnamefont {K.}~\bibnamefont {Kimura}}, \bibinfo {author} {\bibfnamefont
  {S.}~\bibnamefont {Nakatsuji}}, \bibinfo {author} {\bibfnamefont
  {Y.}~\bibnamefont {Qiu}}, \bibinfo {author} {\bibfnamefont {D.~M.}\
  \bibnamefont {Pajerowski}}, \bibinfo {author} {\bibfnamefont {J.~R.~D.}\
  \bibnamefont {Copley}},\ and\ \bibinfo {author} {\bibfnamefont {C.~L.}\
  \bibnamefont {Broholm}},\ }\bibfield  {title} {\bibinfo {title} {{Disordered
  Route to the Coulomb Quantum Spin Liquid: Random Transverse Fields on Spin
  Ice in Pr$_2$Zr$_2$O$_7$}},\ }\href
  {https://doi.org/10.1103/PhysRevLett.118.107206} {\bibfield  {journal}
  {\bibinfo  {journal} {Phys. Rev. Lett.}\ }\textbf {\bibinfo {volume} {118}},\
  \bibinfo {pages} {107206} (\bibinfo {year} {2017})}\BibitemShut {NoStop}%
\bibitem [{\citenamefont {Gao}\ \emph {et~al.}(2019)\citenamefont {Gao},
  \citenamefont {Chen}, \citenamefont {Tam}, \citenamefont {Huang},
  \citenamefont {Sasmal}, \citenamefont {Adroja}, \citenamefont {Ye},
  \citenamefont {Cao}, \citenamefont {Sala}, \citenamefont {Stone},
  \citenamefont {Baines}, \citenamefont {Verezhak}, \citenamefont {Hu},
  \citenamefont {Chung}, \citenamefont {Xu}, \citenamefont {Cheong},
  \citenamefont {Nallaiyan}, \citenamefont {Spagna}, \citenamefont {Maple},
  \citenamefont {Nevidomskyy}, \citenamefont {Morosan}, \citenamefont {Chen},\
  and\ \citenamefont {Dai}}]{Gao2019}%
  \BibitemOpen
  \bibfield  {author} {\bibinfo {author} {\bibfnamefont {B.}~\bibnamefont
  {Gao}}, \bibinfo {author} {\bibfnamefont {T.}~\bibnamefont {Chen}}, \bibinfo
  {author} {\bibfnamefont {D.~W.}\ \bibnamefont {Tam}}, \bibinfo {author}
  {\bibfnamefont {C.~L.}\ \bibnamefont {Huang}}, \bibinfo {author}
  {\bibfnamefont {K.}~\bibnamefont {Sasmal}}, \bibinfo {author} {\bibfnamefont
  {D.~T.}\ \bibnamefont {Adroja}}, \bibinfo {author} {\bibfnamefont
  {F.}~\bibnamefont {Ye}}, \bibinfo {author} {\bibfnamefont {H.}~\bibnamefont
  {Cao}}, \bibinfo {author} {\bibfnamefont {G.}~\bibnamefont {Sala}}, \bibinfo
  {author} {\bibfnamefont {M.~B.}\ \bibnamefont {Stone}}, \bibinfo {author}
  {\bibfnamefont {C.}~\bibnamefont {Baines}}, \bibinfo {author} {\bibfnamefont
  {J.~A.~T.}\ \bibnamefont {Verezhak}}, \bibinfo {author} {\bibfnamefont
  {H.}~\bibnamefont {Hu}}, \bibinfo {author} {\bibfnamefont {J.~H.}\
  \bibnamefont {Chung}}, \bibinfo {author} {\bibfnamefont {X.}~\bibnamefont
  {Xu}}, \bibinfo {author} {\bibfnamefont {S.~W.}\ \bibnamefont {Cheong}},
  \bibinfo {author} {\bibfnamefont {M.}~\bibnamefont {Nallaiyan}}, \bibinfo
  {author} {\bibfnamefont {S.}~\bibnamefont {Spagna}}, \bibinfo {author}
  {\bibfnamefont {M.~B.}\ \bibnamefont {Maple}}, \bibinfo {author}
  {\bibfnamefont {A.~H.}\ \bibnamefont {Nevidomskyy}}, \bibinfo {author}
  {\bibfnamefont {E.}~\bibnamefont {Morosan}}, \bibinfo {author} {\bibfnamefont
  {G.}~\bibnamefont {Chen}},\ and\ \bibinfo {author} {\bibfnamefont
  {P.}~\bibnamefont {Dai}},\ }\bibfield  {title} {\bibinfo {title}
  {{Experimental signatures of a three-dimensional quantum spin liquid in
  effective spin-1/2 Ce$_2$Zr$_2$O$_7$ pyrochlore}},\ }\href
  {https://doi.org/10.1038/s41567-019-0577-6} {\bibfield  {journal} {\bibinfo
  {journal} {Nat. Phys.}\ }\textbf {\bibinfo {volume} {15}},\ \bibinfo {pages}
  {1052} (\bibinfo {year} {2019})}\BibitemShut {NoStop}%
\bibitem [{\citenamefont {Plumb}\ \emph {et~al.}(2019)\citenamefont {Plumb},
  \citenamefont {Changlani}, \citenamefont {Scheie}, \citenamefont {Zhang},
  \citenamefont {Krizan}, \citenamefont {Rodriguez-Rivera}, \citenamefont
  {Qiu}, \citenamefont {Winn}, \citenamefont {Cava},\ and\ \citenamefont
  {Broholm}}]{Plumb2019}%
  \BibitemOpen
  \bibfield  {author} {\bibinfo {author} {\bibfnamefont {K.~W.}\ \bibnamefont
  {Plumb}}, \bibinfo {author} {\bibfnamefont {H.~J.}\ \bibnamefont
  {Changlani}}, \bibinfo {author} {\bibfnamefont {A.}~\bibnamefont {Scheie}},
  \bibinfo {author} {\bibfnamefont {S.}~\bibnamefont {Zhang}}, \bibinfo
  {author} {\bibfnamefont {J.~W.}\ \bibnamefont {Krizan}}, \bibinfo {author}
  {\bibfnamefont {J.~A.}\ \bibnamefont {Rodriguez-Rivera}}, \bibinfo {author}
  {\bibfnamefont {Y.}~\bibnamefont {Qiu}}, \bibinfo {author} {\bibfnamefont
  {B.}~\bibnamefont {Winn}}, \bibinfo {author} {\bibfnamefont {R.~J.}\
  \bibnamefont {Cava}},\ and\ \bibinfo {author} {\bibfnamefont {C.~L.}\
  \bibnamefont {Broholm}},\ }\bibfield  {title} {\bibinfo {title} {{Continuum
  of quantum fluctuations in a three-dimensional $S=1$ Heisenberg magnet}},\
  }\href {https://doi.org/10.1038/s41567-018-0317-3} {\bibfield  {journal}
  {\bibinfo  {journal} {Nat. Phys.}\ }\textbf {\bibinfo {volume} {15}},\
  \bibinfo {pages} {54} (\bibinfo {year} {2019})}\BibitemShut {NoStop}%
\bibitem [{\citenamefont {Okamoto}\ \emph {et~al.}(2007)\citenamefont
  {Okamoto}, \citenamefont {Nohara}, \citenamefont {Aruga-Katori},\ and\
  \citenamefont {Takagi}}]{Okamoto2007}%
  \BibitemOpen
  \bibfield  {author} {\bibinfo {author} {\bibfnamefont {Y.}~\bibnamefont
  {Okamoto}}, \bibinfo {author} {\bibfnamefont {M.}~\bibnamefont {Nohara}},
  \bibinfo {author} {\bibfnamefont {H.}~\bibnamefont {Aruga-Katori}},\ and\
  \bibinfo {author} {\bibfnamefont {H.}~\bibnamefont {Takagi}},\ }\bibfield
  {title} {\bibinfo {title} {{Spin-Liquid State in the S = 1/2 Hyperkagome
  Antiferromagnet Na$_4$Ir$_3$O$_8$}},\ }\href
  {https://doi.org/10.1103/PhysRevLett.99.137207} {\bibfield  {journal}
  {\bibinfo  {journal} {Phys. Rev. Lett.}\ }\textbf {\bibinfo {volume} {99}},\
  \bibinfo {pages} {137207} (\bibinfo {year} {2007})}\BibitemShut {NoStop}%
\bibitem [{\citenamefont {Chillal}\ \emph {et~al.}(2020)\citenamefont
  {Chillal}, \citenamefont {Iqbal}, \citenamefont {Jeschke}, \citenamefont
  {Rodriguez-Rivera}, \citenamefont {Bewley}, \citenamefont {Manuel},
  \citenamefont {Khalyavin}, \citenamefont {Steffens}, \citenamefont {Thomale},
  \citenamefont {Nazmul~Islam}, \citenamefont {Reuther},\ and\ \citenamefont
  {Lake}}]{Chillal2020}%
  \BibitemOpen
  \bibfield  {author} {\bibinfo {author} {\bibfnamefont {S.}~\bibnamefont
  {Chillal}}, \bibinfo {author} {\bibfnamefont {Y.}~\bibnamefont {Iqbal}},
  \bibinfo {author} {\bibfnamefont {H.~O.}\ \bibnamefont {Jeschke}}, \bibinfo
  {author} {\bibfnamefont {J.~A.}\ \bibnamefont {Rodriguez-Rivera}}, \bibinfo
  {author} {\bibfnamefont {R.}~\bibnamefont {Bewley}}, \bibinfo {author}
  {\bibfnamefont {P.}~\bibnamefont {Manuel}}, \bibinfo {author} {\bibfnamefont
  {D.}~\bibnamefont {Khalyavin}}, \bibinfo {author} {\bibfnamefont
  {P.}~\bibnamefont {Steffens}}, \bibinfo {author} {\bibfnamefont
  {P.}~\bibnamefont {Thomale}}, \bibinfo {author} {\bibfnamefont {A.~T.~M.}\
  \bibnamefont {Nazmul~Islam}}, \bibinfo {author} {\bibfnamefont
  {J.}~\bibnamefont {Reuther}},\ and\ \bibinfo {author} {\bibfnamefont
  {B.}~\bibnamefont {Lake}},\ }\bibfield  {title} {\bibinfo {title} {{Evidence
  for a three-dimensional quantum spin liquid in PbCuTe$_2$O$_6$}},\ }\href
  {https://doi.org/10.1038/s41467-020-15594-1} {\bibfield  {journal} {\bibinfo
  {journal} {Nat. Comm.}\ }\textbf {\bibinfo {volume} {11}},\ \bibinfo {pages}
  {2348} (\bibinfo {year} {2020})}\BibitemShut {NoStop}%
\bibitem [{\citenamefont {Pfleiderer}\ \emph {et~al.}(1997)\citenamefont
  {Pfleiderer}, \citenamefont {McMullan}, \citenamefont {Julian},\ and\
  \citenamefont {Lonzarich}}]{Pfleiderer1997}%
  \BibitemOpen
  \bibfield  {author} {\bibinfo {author} {\bibfnamefont {C.}~\bibnamefont
  {Pfleiderer}}, \bibinfo {author} {\bibfnamefont {G.~J.}\ \bibnamefont
  {McMullan}}, \bibinfo {author} {\bibfnamefont {S.~R.}\ \bibnamefont
  {Julian}},\ and\ \bibinfo {author} {\bibfnamefont {G.~G.}\ \bibnamefont
  {Lonzarich}},\ }\bibfield  {title} {\bibinfo {title} {{Magnetic quantum phase
  transition in MnSi under hydrostatic pressure}},\ }\href
  {https://doi.org/10.1103/PhysRevB.55.8330} {\bibfield  {journal} {\bibinfo
  {journal} {Phys. Rev. B}\ }\textbf {\bibinfo {volume} {55}},\ \bibinfo
  {pages} {8330} (\bibinfo {year} {1997})}\BibitemShut {NoStop}%
\bibitem [{\citenamefont {Pfleiderer}\ \emph {et~al.}(2001)\citenamefont
  {Pfleiderer}, \citenamefont {Julian},\ and\ \citenamefont
  {Lonzarich}}]{Pfleiderer2001}%
  \BibitemOpen
  \bibfield  {author} {\bibinfo {author} {\bibfnamefont {C.}~\bibnamefont
  {Pfleiderer}}, \bibinfo {author} {\bibfnamefont {S.~R.}\ \bibnamefont
  {Julian}},\ and\ \bibinfo {author} {\bibfnamefont {G.~G.}\ \bibnamefont
  {Lonzarich}},\ }\bibfield  {title} {\bibinfo {title} {{Non-Fermi-liquid
  nature of the normal state of itinerant-electron ferromagnets}},\ }\href
  {https://doi.org/10.1038/35106527} {\bibfield  {journal} {\bibinfo  {journal}
  {Nature (London)}\ }\textbf {\bibinfo {volume} {414}},\ \bibinfo {pages}
  {427} (\bibinfo {year} {2001})}\BibitemShut {NoStop}%
\bibitem [{\citenamefont {Hopkinson}\ and\ \citenamefont
  {Kee}(2006)}]{Hopkinson2006}%
  \BibitemOpen
  \bibfield  {author} {\bibinfo {author} {\bibfnamefont {J.~M.}\ \bibnamefont
  {Hopkinson}}\ and\ \bibinfo {author} {\bibfnamefont {H.-Y.}\ \bibnamefont
  {Kee}},\ }\bibfield  {title} {\bibinfo {title} {{Geometric frustration
  inherent to the trillium lattice, a sublattice of the B20 structure}},\
  }\href {https://doi.org/10.1103/PhysRevB.74.224441} {\bibfield  {journal}
  {\bibinfo  {journal} {Phys. Rev. B}\ }\textbf {\bibinfo {volume} {74}},\
  \bibinfo {pages} {224441} (\bibinfo {year} {2006})}\BibitemShut {NoStop}%
\bibitem [{\citenamefont {Isakov}\ \emph {et~al.}(2008)\citenamefont {Isakov},
  \citenamefont {Hopkinson},\ and\ \citenamefont {Kee}}]{Isakov2008}%
  \BibitemOpen
  \bibfield  {author} {\bibinfo {author} {\bibfnamefont {S.~V.}\ \bibnamefont
  {Isakov}}, \bibinfo {author} {\bibfnamefont {J.~M.}\ \bibnamefont
  {Hopkinson}},\ and\ \bibinfo {author} {\bibfnamefont {H.~Y.}\ \bibnamefont
  {Kee}},\ }\bibfield  {title} {\bibinfo {title} {{Fate of partial order on
  trillium and distorted windmill lattices}},\ }\href
  {https://doi.org/10.1103/PhysRevB.78.014404} {\bibfield  {journal} {\bibinfo
  {journal} {Phys. Rev. B}\ }\textbf {\bibinfo {volume} {78}},\ \bibinfo
  {pages} {014404} (\bibinfo {year} {2008})}\BibitemShut {NoStop}%
\bibitem [{SM()}]{SM}%
  \BibitemOpen
  \href@noop {} {}\bibinfo {note} {See Supplemental Material at
  http://link.aps.org/supplemental/10.1103/PhysRevLett.XXX for additional
  information on sample preparation, experimental and computational methods,
  powder diffraction, crystal structure data, magnetization, specific heat,
  neutron scattering, muon spin relaxation experiments and PFFRG
  calculations.}\BibitemShut {Stop}%
\bibitem [{\citenamefont {Ramirez}\ \emph {et~al.}(2000)\citenamefont
  {Ramirez}, \citenamefont {Hessen},\ and\ \citenamefont
  {Winklemann}}]{Ramirez2000}%
  \BibitemOpen
  \bibfield  {author} {\bibinfo {author} {\bibfnamefont {A.~P.}\ \bibnamefont
  {Ramirez}}, \bibinfo {author} {\bibfnamefont {B.}~\bibnamefont {Hessen}},\
  and\ \bibinfo {author} {\bibfnamefont {M.}~\bibnamefont {Winklemann}},\
  }\bibfield  {title} {\bibinfo {title} {{Entropy Balance and Evidence for
  Local Spin Singlets in a Kagom\'e-Like Magnet}},\ }\href
  {https://doi.org/10.1103/PhysRevLett.84.2957} {\bibfield  {journal} {\bibinfo
   {journal} {Phys. Rev. Lett.}\ }\textbf {\bibinfo {volume} {84}},\ \bibinfo
  {pages} {2957} (\bibinfo {year} {2000})}\BibitemShut {NoStop}%
\bibitem [{\citenamefont {Nakatsuji}\ \emph {et~al.}(2005)\citenamefont
  {Nakatsuji}, \citenamefont {Nambu}, \citenamefont {Tonomura}, \citenamefont
  {Sakai}, \citenamefont {Broholm}, \citenamefont {Tsunetsugu}, \citenamefont
  {Qiu},\ and\ \citenamefont {Maeno}}]{Nakatsuji2005}%
  \BibitemOpen
  \bibfield  {author} {\bibinfo {author} {\bibfnamefont {S.}~\bibnamefont
  {Nakatsuji}}, \bibinfo {author} {\bibfnamefont {Y.}~\bibnamefont {Nambu}},
  \bibinfo {author} {\bibfnamefont {H.}~\bibnamefont {Tonomura}}, \bibinfo
  {author} {\bibfnamefont {S.}~\bibnamefont {Sakai}, \bibfnamefont {O.~Jonas}},
  \bibinfo {author} {\bibfnamefont {C.}~\bibnamefont {Broholm}}, \bibinfo
  {author} {\bibfnamefont {H.}~\bibnamefont {Tsunetsugu}}, \bibinfo {author}
  {\bibfnamefont {Y.}~\bibnamefont {Qiu}},\ and\ \bibinfo {author}
  {\bibfnamefont {Y.}~\bibnamefont {Maeno}},\ }\bibfield  {title} {\bibinfo
  {title} {{Spin Disorder on a Triangular Lattice}},\ }\href
  {https://doi.org/10.1126/science.1114727} {\bibfield  {journal} {\bibinfo
  {journal} {Science}\ }\textbf {\bibinfo {volume} {309}},\ \bibinfo {pages}
  {1697} (\bibinfo {year} {2005})}\BibitemShut {NoStop}%
\bibitem [{\citenamefont {Silverstein}\ \emph {et~al.}(2014)\citenamefont
  {Silverstein}, \citenamefont {Fritsch}, \citenamefont {Flicker},
  \citenamefont {Hallas}, \citenamefont {Gardner}, \citenamefont {Qiu},
  \citenamefont {Ehlers}, \citenamefont {Savici}, \citenamefont {Yamani},
  \citenamefont {Ross}, \citenamefont {Gaulin}, \citenamefont {Gingras},
  \citenamefont {Paddison}, \citenamefont {Foyevtsova}, \citenamefont
  {Valenti}, \citenamefont {Hawthorne}, \citenamefont {Wiebe},\ and\
  \citenamefont {Zhou}}]{Silverstein2014}%
  \BibitemOpen
  \bibfield  {author} {\bibinfo {author} {\bibfnamefont {H.~J.}\ \bibnamefont
  {Silverstein}}, \bibinfo {author} {\bibfnamefont {K.}~\bibnamefont
  {Fritsch}}, \bibinfo {author} {\bibfnamefont {F.}~\bibnamefont {Flicker}},
  \bibinfo {author} {\bibfnamefont {A.~M.}\ \bibnamefont {Hallas}}, \bibinfo
  {author} {\bibfnamefont {J.~S.}\ \bibnamefont {Gardner}}, \bibinfo {author}
  {\bibfnamefont {Y.}~\bibnamefont {Qiu}}, \bibinfo {author} {\bibfnamefont
  {G.}~\bibnamefont {Ehlers}}, \bibinfo {author} {\bibfnamefont {A.~T.}\
  \bibnamefont {Savici}}, \bibinfo {author} {\bibfnamefont {Z.}~\bibnamefont
  {Yamani}}, \bibinfo {author} {\bibfnamefont {K.~A.}\ \bibnamefont {Ross}},
  \bibinfo {author} {\bibfnamefont {B.~D.}\ \bibnamefont {Gaulin}}, \bibinfo
  {author} {\bibfnamefont {M.~J.~P.}\ \bibnamefont {Gingras}}, \bibinfo
  {author} {\bibfnamefont {J.~A.~M.}\ \bibnamefont {Paddison}}, \bibinfo
  {author} {\bibfnamefont {K.}~\bibnamefont {Foyevtsova}}, \bibinfo {author}
  {\bibfnamefont {R.}~\bibnamefont {Valenti}}, \bibinfo {author} {\bibfnamefont
  {F.}~\bibnamefont {Hawthorne}}, \bibinfo {author} {\bibfnamefont {C.~R.}\
  \bibnamefont {Wiebe}},\ and\ \bibinfo {author} {\bibfnamefont {H.~D.}\
  \bibnamefont {Zhou}},\ }\bibfield  {title} {\bibinfo {title} {{Liquidlike
  correlations in single-crystalline Y$_2$Mo$_2$O$_7$: An unconventional spin
  glass}},\ }\href {https://doi.org/10.1103/PhysRevB.89.054433} {\bibfield
  {journal} {\bibinfo  {journal} {Phys. Rev. B}\ }\textbf {\bibinfo {volume}
  {89}},\ \bibinfo {pages} {054433} (\bibinfo {year} {2014})}\BibitemShut
  {NoStop}%
\bibitem [{\citenamefont {Boerio-Goates}\ \emph {et~al.}(1990)\citenamefont
  {Boerio-Goates}, \citenamefont {Artman},\ and\ \citenamefont
  {Woodfield}}]{Boerio1990}%
  \BibitemOpen
  \bibfield  {author} {\bibinfo {author} {\bibfnamefont {J.}~\bibnamefont
  {Boerio-Goates}}, \bibinfo {author} {\bibfnamefont {J.~I.}\ \bibnamefont
  {Artman}},\ and\ \bibinfo {author} {\bibfnamefont {B.~F.}\ \bibnamefont
  {Woodfield}},\ }\bibfield  {title} {\bibinfo {title} {{Heat Capacity Studies
  of Phase Transitions in Langbeinites II. K$_2$Mg$_2$(SO$_4$)$_3$}},\ }\href
  {https://doi.org/10.1007/BF00199670} {\bibfield  {journal} {\bibinfo
  {journal} {Phys. Chem. Minerals}\ }\textbf {\bibinfo {volume} {17}},\
  \bibinfo {pages} {173} (\bibinfo {year} {1990})}\BibitemShut {NoStop}%
\bibitem [{\citenamefont {Xu}\ \emph {et~al.}(2016)\citenamefont {Xu},
  \citenamefont {Zhang}, \citenamefont {Li}, \citenamefont {Yu}, \citenamefont
  {Hong}, \citenamefont {Zhang},\ and\ \citenamefont {Li}}]{Xu2016}%
  \BibitemOpen
  \bibfield  {author} {\bibinfo {author} {\bibfnamefont {Y.}~\bibnamefont
  {Xu}}, \bibinfo {author} {\bibfnamefont {J.}~\bibnamefont {Zhang}}, \bibinfo
  {author} {\bibfnamefont {Y.~S.}\ \bibnamefont {Li}}, \bibinfo {author}
  {\bibfnamefont {Y.~J.}\ \bibnamefont {Yu}}, \bibinfo {author} {\bibfnamefont
  {X.~C.}\ \bibnamefont {Hong}}, \bibinfo {author} {\bibfnamefont {Q.~M.}\
  \bibnamefont {Zhang}},\ and\ \bibinfo {author} {\bibfnamefont {S.~Y.}\
  \bibnamefont {Li}},\ }\bibfield  {title} {\bibinfo {title} {{Absence of
  Magnetic Thermal Conductivity in the Quantum Spin-Liquid Candidate
  YbMgGaO$_4$}},\ }\href {https://doi.org/10.1103/PhysRevLett.117.267202}
  {\bibfield  {journal} {\bibinfo  {journal} {Phys. Rev. Lett.}\ }\textbf
  {\bibinfo {volume} {117}},\ \bibinfo {pages} {267202} (\bibinfo {year}
  {2016})}\BibitemShut {NoStop}%
\bibitem [{\citenamefont {Mendels}\ \emph {et~al.}(2007)\citenamefont
  {Mendels}, \citenamefont {Bert}, \citenamefont {de~Vries}, \citenamefont
  {Olariu}, \citenamefont {Harrison}, \citenamefont {Duc}, \citenamefont
  {Trombe}, \citenamefont {Lord}, \citenamefont {Amato},\ and\ \citenamefont
  {Baines}}]{Mendels2007}%
  \BibitemOpen
  \bibfield  {author} {\bibinfo {author} {\bibfnamefont {P.}~\bibnamefont
  {Mendels}}, \bibinfo {author} {\bibfnamefont {F.}~\bibnamefont {Bert}},
  \bibinfo {author} {\bibfnamefont {M.~A.}\ \bibnamefont {de~Vries}}, \bibinfo
  {author} {\bibfnamefont {A.}~\bibnamefont {Olariu}}, \bibinfo {author}
  {\bibfnamefont {A.}~\bibnamefont {Harrison}}, \bibinfo {author}
  {\bibfnamefont {F.}~\bibnamefont {Duc}}, \bibinfo {author} {\bibfnamefont
  {J.~C.}\ \bibnamefont {Trombe}}, \bibinfo {author} {\bibfnamefont {J.~S.}\
  \bibnamefont {Lord}}, \bibinfo {author} {\bibfnamefont {A.}~\bibnamefont
  {Amato}},\ and\ \bibinfo {author} {\bibfnamefont {C.}~\bibnamefont
  {Baines}},\ }\bibfield  {title} {\bibinfo {title} {{Quantum Magnetism in the
  Paratacamite Family: Towards an Ideal Kagome Lattice}},\ }\href
  {https://doi.org/10.1103/PhysRevLett.98.077204} {\bibfield  {journal}
  {\bibinfo  {journal} {Phys. Rev. Lett.}\ }\textbf {\bibinfo {volume} {98}},\
  \bibinfo {pages} {077204} (\bibinfo {year} {2007})}\BibitemShut {NoStop}%
\bibitem [{\citenamefont {Balz}\ \emph {et~al.}(2016)\citenamefont {Balz},
  \citenamefont {Lake}, \citenamefont {Reuther}, \citenamefont {Luetkens},
  \citenamefont {Sch{\"o}nemann}, \citenamefont {Herrmannsd{\"o}rfer},
  \citenamefont {Singh}, \citenamefont {Nazmul~Islam}, \citenamefont {Wheeler},
  \citenamefont {Rodriguez-Rivera}, \citenamefont {Guidi}, \citenamefont
  {Simeoni}, \citenamefont {Baines},\ and\ \citenamefont {Ryll}}]{Balz16}%
  \BibitemOpen
  \bibfield  {author} {\bibinfo {author} {\bibfnamefont {C.}~\bibnamefont
  {Balz}}, \bibinfo {author} {\bibfnamefont {B.}~\bibnamefont {Lake}}, \bibinfo
  {author} {\bibfnamefont {J.}~\bibnamefont {Reuther}}, \bibinfo {author}
  {\bibfnamefont {H.}~\bibnamefont {Luetkens}}, \bibinfo {author}
  {\bibfnamefont {R.}~\bibnamefont {Sch{\"o}nemann}}, \bibinfo {author}
  {\bibfnamefont {T.}~\bibnamefont {Herrmannsd{\"o}rfer}}, \bibinfo {author}
  {\bibfnamefont {Y.}~\bibnamefont {Singh}}, \bibinfo {author} {\bibfnamefont
  {A.~T.~M.}\ \bibnamefont {Nazmul~Islam}}, \bibinfo {author} {\bibfnamefont
  {E.~M.}\ \bibnamefont {Wheeler}}, \bibinfo {author} {\bibfnamefont {J.~A.}\
  \bibnamefont {Rodriguez-Rivera}}, \bibinfo {author} {\bibfnamefont
  {T.}~\bibnamefont {Guidi}}, \bibinfo {author} {\bibfnamefont {G.~G.}\
  \bibnamefont {Simeoni}}, \bibinfo {author} {\bibfnamefont {C.}~\bibnamefont
  {Baines}},\ and\ \bibinfo {author} {\bibfnamefont {H.}~\bibnamefont {Ryll}},\
  }\bibfield  {title} {\bibinfo {title} {Physical realization of a quantum spin
  liquid based on a complex frustration mechanism},\ }\href
  {http://dx.doi.org/10.1038/nphys3826} {\bibfield  {journal} {\bibinfo
  {journal} {Nat. Phys.}\ }\textbf {\bibinfo {volume} {12}},\ \bibinfo {pages}
  {942} (\bibinfo {year} {2016})}\BibitemShut {NoStop}%
\bibitem [{\citenamefont {Fujihala}\ \emph {et~al.}(2020)\citenamefont
  {Fujihala}, \citenamefont {Morita}, \citenamefont {Mole}, \citenamefont
  {Mitsuda}, \citenamefont {Tohyama}, \citenamefont {Yano}, \citenamefont {Yu},
  \citenamefont {Sota}, \citenamefont {Kuwai}, \citenamefont {Koda},
  \citenamefont {Okabe}, \citenamefont {Lee}, \citenamefont {Itoh},
  \citenamefont {Hawai}, \citenamefont {masuda}, \citenamefont {Sagayama},
  \citenamefont {Matsuo}, \citenamefont {Kindo}, \citenamefont
  {Ohira-Kawamura},\ and\ \citenamefont {Nakajima}}]{Fujihala2020}%
  \BibitemOpen
  \bibfield  {author} {\bibinfo {author} {\bibfnamefont {M.}~\bibnamefont
  {Fujihala}}, \bibinfo {author} {\bibfnamefont {K.}~\bibnamefont {Morita}},
  \bibinfo {author} {\bibfnamefont {R.}~\bibnamefont {Mole}}, \bibinfo {author}
  {\bibfnamefont {S.}~\bibnamefont {Mitsuda}}, \bibinfo {author} {\bibfnamefont
  {T.}~\bibnamefont {Tohyama}}, \bibinfo {author} {\bibfnamefont
  {S.}~\bibnamefont {Yano}}, \bibinfo {author} {\bibfnamefont {D.}~\bibnamefont
  {Yu}}, \bibinfo {author} {\bibfnamefont {S.}~\bibnamefont {Sota}}, \bibinfo
  {author} {\bibfnamefont {T.}~\bibnamefont {Kuwai}}, \bibinfo {author}
  {\bibfnamefont {A.}~\bibnamefont {Koda}}, \bibinfo {author} {\bibfnamefont
  {H.}~\bibnamefont {Okabe}}, \bibinfo {author} {\bibfnamefont
  {H.}~\bibnamefont {Lee}}, \bibinfo {author} {\bibfnamefont {S.}~\bibnamefont
  {Itoh}}, \bibinfo {author} {\bibfnamefont {T.}~\bibnamefont {Hawai}},
  \bibinfo {author} {\bibfnamefont {T.}~\bibnamefont {masuda}}, \bibinfo
  {author} {\bibfnamefont {H.}~\bibnamefont {Sagayama}}, \bibinfo {author}
  {\bibfnamefont {A.}~\bibnamefont {Matsuo}}, \bibinfo {author} {\bibfnamefont
  {K.}~\bibnamefont {Kindo}}, \bibinfo {author} {\bibfnamefont
  {S.}~\bibnamefont {Ohira-Kawamura}},\ and\ \bibinfo {author} {\bibfnamefont
  {K.}~\bibnamefont {Nakajima}},\ }\bibfield  {title} {\bibinfo {title}
  {{Gapless spin liquid in a square-kagome lattice antiferromagnet}},\ }\href
  {https://doi.org/10.1038/s41467-020-17235-z} {\bibfield  {journal} {\bibinfo
  {journal} {Nat. Comm.}\ }\textbf {\bibinfo {volume} {11}},\ \bibinfo {pages}
  {3429} (\bibinfo {year} {2020})}\BibitemShut {NoStop}%
\bibitem [{\citenamefont {Uemura}\ \emph {et~al.}(1994)\citenamefont {Uemura},
  \citenamefont {Keren}, \citenamefont {Kojima}, \citenamefont {Le},
  \citenamefont {Luke}, \citenamefont {Wu}, \citenamefont {Ajiro},
  \citenamefont {Asano}, \citenamefont {Kuriyama}, \citenamefont {Mekata},
  \citenamefont {Kikuchi},\ and\ \citenamefont {Kakurai}}]{Uemura1994}%
  \BibitemOpen
  \bibfield  {author} {\bibinfo {author} {\bibfnamefont {Y.~J.}\ \bibnamefont
  {Uemura}}, \bibinfo {author} {\bibfnamefont {A.}~\bibnamefont {Keren}},
  \bibinfo {author} {\bibfnamefont {K.}~\bibnamefont {Kojima}}, \bibinfo
  {author} {\bibfnamefont {L.~P.}\ \bibnamefont {Le}}, \bibinfo {author}
  {\bibfnamefont {G.~M.}\ \bibnamefont {Luke}}, \bibinfo {author}
  {\bibfnamefont {W.~D.}\ \bibnamefont {Wu}}, \bibinfo {author} {\bibfnamefont
  {Y.}~\bibnamefont {Ajiro}}, \bibinfo {author} {\bibfnamefont
  {T.}~\bibnamefont {Asano}}, \bibinfo {author} {\bibfnamefont
  {Y.}~\bibnamefont {Kuriyama}}, \bibinfo {author} {\bibfnamefont
  {M.}~\bibnamefont {Mekata}}, \bibinfo {author} {\bibfnamefont
  {H.}~\bibnamefont {Kikuchi}},\ and\ \bibinfo {author} {\bibfnamefont
  {K.}~\bibnamefont {Kakurai}},\ }\bibfield  {title} {\bibinfo {title} {{Spin
  Fluctuations in Frustrated Kagome Lattice System SrCr$_8$Ga$_4$O$_{19}$
  Studied by Muon Spin Relaxation}},\ }\href
  {https://doi.org/10.1103/PhysRevLett.73.3306} {\bibfield  {journal} {\bibinfo
   {journal} {Phys. Rev. Lett.}\ }\textbf {\bibinfo {volume} {73}},\ \bibinfo
  {pages} {3306} (\bibinfo {year} {1994})}\BibitemShut {NoStop}%
\bibitem [{\citenamefont {R\"uegg}\ \emph {et~al.}(2008)\citenamefont
  {R\"uegg}, \citenamefont {Normand}, \citenamefont {Matsumoto}, \citenamefont
  {Furrer}, \citenamefont {McMorrow}, \citenamefont {Kr\"amer}, \citenamefont
  {G\"udel}, \citenamefont {Gvasaliya}, \citenamefont {Mutka},\ and\
  \citenamefont {Boehm}}]{Ruegg2008}%
  \BibitemOpen
  \bibfield  {author} {\bibinfo {author} {\bibfnamefont {C.}~\bibnamefont
  {R\"uegg}}, \bibinfo {author} {\bibfnamefont {B.}~\bibnamefont {Normand}},
  \bibinfo {author} {\bibfnamefont {M.}~\bibnamefont {Matsumoto}}, \bibinfo
  {author} {\bibfnamefont {A.}~\bibnamefont {Furrer}}, \bibinfo {author}
  {\bibfnamefont {D.~F.}\ \bibnamefont {McMorrow}}, \bibinfo {author}
  {\bibfnamefont {K.~W.}\ \bibnamefont {Kr\"amer}}, \bibinfo {author}
  {\bibfnamefont {H.~U.}\ \bibnamefont {G\"udel}}, \bibinfo {author}
  {\bibfnamefont {S.~N.}\ \bibnamefont {Gvasaliya}}, \bibinfo {author}
  {\bibfnamefont {H.}~\bibnamefont {Mutka}},\ and\ \bibinfo {author}
  {\bibfnamefont {M.}~\bibnamefont {Boehm}},\ }\bibfield  {title} {\bibinfo
  {title} {{Quantum Magnets under Pressure: Controlling Elementary Excitations
  in TlCuCl$_3$}},\ }\href {https://doi.org/10.1103/PhysRevLett.100.205701}
  {\bibfield  {journal} {\bibinfo  {journal} {Phys. Rev. Lett.}\ }\textbf
  {\bibinfo {volume} {100}},\ \bibinfo {pages} {205701} (\bibinfo {year}
  {2008})}\BibitemShut {NoStop}%
\bibitem [{\citenamefont {M\"{u}hlbauer}\ \emph {et~al.}(2009)\citenamefont
  {M\"{u}hlbauer}, \citenamefont {Binz}, \citenamefont {Jonietz}, \citenamefont
  {Pfleiderer}, \citenamefont {Rosch}, \citenamefont {Neubauer}, \citenamefont
  {Georgii},\ and\ \citenamefont {B\"{o}ni}}]{Muhlbauer2009}%
  \BibitemOpen
  \bibfield  {author} {\bibinfo {author} {\bibfnamefont {S.}~\bibnamefont
  {M\"{u}hlbauer}}, \bibinfo {author} {\bibfnamefont {B.}~\bibnamefont {Binz}},
  \bibinfo {author} {\bibfnamefont {F.}~\bibnamefont {Jonietz}}, \bibinfo
  {author} {\bibfnamefont {C.}~\bibnamefont {Pfleiderer}}, \bibinfo {author}
  {\bibfnamefont {A.}~\bibnamefont {Rosch}}, \bibinfo {author} {\bibfnamefont
  {A.}~\bibnamefont {Neubauer}}, \bibinfo {author} {\bibfnamefont
  {R.}~\bibnamefont {Georgii}},\ and\ \bibinfo {author} {\bibfnamefont
  {P.}~\bibnamefont {B\"{o}ni}},\ }\bibfield  {title} {\bibinfo {title}
  {Skyrmion lattice in a chiral magnet},\ }\href
  {https://doi.org/10.1126/science.1166767} {\bibfield  {journal} {\bibinfo
  {journal} {Science}\ }\textbf {\bibinfo {volume} {323}},\ \bibinfo {pages}
  {915} (\bibinfo {year} {2009})}\BibitemShut {NoStop}%
\bibitem [{\citenamefont {Seki}\ \emph {et~al.}(2012)\citenamefont {Seki},
  \citenamefont {Yu}, \citenamefont {Ishiwata},\ and\ \citenamefont
  {Tokura}}]{Seki2012}%
  \BibitemOpen
  \bibfield  {author} {\bibinfo {author} {\bibfnamefont {S.}~\bibnamefont
  {Seki}}, \bibinfo {author} {\bibfnamefont {X.~Z.}\ \bibnamefont {Yu}},
  \bibinfo {author} {\bibfnamefont {S.}~\bibnamefont {Ishiwata}},\ and\
  \bibinfo {author} {\bibfnamefont {Y.}~\bibnamefont {Tokura}},\ }\bibfield
  {title} {\bibinfo {title} {Observation of skyrmions in a multiferroic
  material},\ }\href {https://doi.org/10.1126/science.1214143} {\bibfield
  {journal} {\bibinfo  {journal} {Science}\ }\textbf {\bibinfo {volume}
  {336}},\ \bibinfo {pages} {198} (\bibinfo {year} {2012})}\BibitemShut
  {NoStop}%
\bibitem [{\citenamefont {Kitaev}(2006)}]{Kitaev2006}%
  \BibitemOpen
  \bibfield  {author} {\bibinfo {author} {\bibfnamefont {A.}~\bibnamefont
  {Kitaev}},\ }\bibfield  {title} {\bibinfo {title} {{Anyons in an exactly
  solved model and beyond}},\ }\href
  {https://doi.org/10.1016/j.aop.2005.10.005} {\bibfield  {journal} {\bibinfo
  {journal} {Ann. Phys.}\ }\textbf {\bibinfo {volume} {321}},\ \bibinfo {pages}
  {2} (\bibinfo {year} {2006})}\BibitemShut {NoStop}%
\bibitem [{\citenamefont {Koepernik}\ and\ \citenamefont
  {Eschrig}(1999)}]{Koepernik1999}%
  \BibitemOpen
  \bibfield  {author} {\bibinfo {author} {\bibfnamefont {K.}~\bibnamefont
  {Koepernik}}\ and\ \bibinfo {author} {\bibfnamefont {H.}~\bibnamefont
  {Eschrig}},\ }\bibfield  {title} {\bibinfo {title} {Full-potential
  nonorthogonal local-orbital minimum-basis band-structure scheme},\ }\href
  {https://doi.org/10.1103/PhysRevB.59.1743} {\bibfield  {journal} {\bibinfo
  {journal} {Phys. Rev. B}\ }\textbf {\bibinfo {volume} {59}},\ \bibinfo
  {pages} {1743} (\bibinfo {year} {1999})}\BibitemShut {NoStop}%
\bibitem [{\citenamefont {Perdew}\ \emph {et~al.}(1996)\citenamefont {Perdew},
  \citenamefont {Burke},\ and\ \citenamefont {Ernzerhof}}]{Perdew1996}%
  \BibitemOpen
  \bibfield  {author} {\bibinfo {author} {\bibfnamefont {J.~P.}\ \bibnamefont
  {Perdew}}, \bibinfo {author} {\bibfnamefont {K.}~\bibnamefont {Burke}},\ and\
  \bibinfo {author} {\bibfnamefont {M.}~\bibnamefont {Ernzerhof}},\ }\bibfield
  {title} {\bibinfo {title} {Generalized gradient approximation made simple},\
  }\href {https://doi.org/10.1103/PhysRevLett.77.3865} {\bibfield  {journal}
  {\bibinfo  {journal} {Phys. Rev. Lett.}\ }\textbf {\bibinfo {volume} {77}},\
  \bibinfo {pages} {3865} (\bibinfo {year} {1996})}\BibitemShut {NoStop}%
\bibitem [{\citenamefont {Liechtenstein}\ \emph {et~al.}(1995)\citenamefont
  {Liechtenstein}, \citenamefont {Anisimov},\ and\ \citenamefont
  {Zaanen}}]{Liechtenstein1995}%
  \BibitemOpen
  \bibfield  {author} {\bibinfo {author} {\bibfnamefont {A.~I.}\ \bibnamefont
  {Liechtenstein}}, \bibinfo {author} {\bibfnamefont {V.~I.}\ \bibnamefont
  {Anisimov}},\ and\ \bibinfo {author} {\bibfnamefont {J.}~\bibnamefont
  {Zaanen}},\ }\bibfield  {title} {\bibinfo {title} {Density-functional theory
  and strong interactions: Orbital ordering in {Mott-Hubbard} insulators},\
  }\href {https://doi.org/10.1103/PhysRevB.52.R5467} {\bibfield  {journal}
  {\bibinfo  {journal} {Phys. Rev. B}\ }\textbf {\bibinfo {volume} {52}},\
  \bibinfo {pages} {R5467} (\bibinfo {year} {1995})}\BibitemShut {NoStop}%
\bibitem [{\citenamefont {Jeschke}\ \emph {et~al.}(2015)\citenamefont
  {Jeschke}, \citenamefont {Salvat-Pujol}, \citenamefont {Gati}, \citenamefont
  {Hoang}, \citenamefont {Wolf}, \citenamefont {Lang}, \citenamefont
  {Schlueter},\ and\ \citenamefont {Valent\'{\i}}}]{Jeschke2015}%
  \BibitemOpen
  \bibfield  {author} {\bibinfo {author} {\bibfnamefont {H.~O.}\ \bibnamefont
  {Jeschke}}, \bibinfo {author} {\bibfnamefont {F.}~\bibnamefont
  {Salvat-Pujol}}, \bibinfo {author} {\bibfnamefont {E.}~\bibnamefont {Gati}},
  \bibinfo {author} {\bibfnamefont {N.~H.}\ \bibnamefont {Hoang}}, \bibinfo
  {author} {\bibfnamefont {B.}~\bibnamefont {Wolf}}, \bibinfo {author}
  {\bibfnamefont {M.}~\bibnamefont {Lang}}, \bibinfo {author} {\bibfnamefont
  {J.~A.}\ \bibnamefont {Schlueter}},\ and\ \bibinfo {author} {\bibfnamefont
  {R.}~\bibnamefont {Valent\'{\i}}},\ }\bibfield  {title} {\bibinfo {title}
  {Barlowite as a canted antiferromagnet: Theory and experiment},\ }\href
  {https://doi.org/10.1103/PhysRevB.92.094417} {\bibfield  {journal} {\bibinfo
  {journal} {Phys. Rev. B}\ }\textbf {\bibinfo {volume} {92}},\ \bibinfo
  {pages} {094417} (\bibinfo {year} {2015})}\BibitemShut {NoStop}%
\bibitem [{\citenamefont {Jeschke}\ \emph {et~al.}(2019)\citenamefont
  {Jeschke}, \citenamefont {Nakano},\ and\ \citenamefont
  {Sakai}}]{Jeschke2019}%
  \BibitemOpen
  \bibfield  {author} {\bibinfo {author} {\bibfnamefont {H.~O.}\ \bibnamefont
  {Jeschke}}, \bibinfo {author} {\bibfnamefont {H.}~\bibnamefont {Nakano}},\
  and\ \bibinfo {author} {\bibfnamefont {T.}~\bibnamefont {Sakai}},\ }\bibfield
   {title} {\bibinfo {title} {From kagome strip to kagome lattice: Realizations
  of frustrated {$S=\frac{1}{2}$} antiferromagnets in {Ti(III)} fluorides},\
  }\href {https://doi.org/10.1103/PhysRevB.99.140410} {\bibfield  {journal}
  {\bibinfo  {journal} {Phys. Rev. B}\ }\textbf {\bibinfo {volume} {99}},\
  \bibinfo {pages} {140410} (\bibinfo {year} {2019})}\BibitemShut {NoStop}%
\bibitem [{\citenamefont {Mizokawa}\ and\ \citenamefont
  {Fujimori}(1996)}]{Mizokawa1996}%
  \BibitemOpen
  \bibfield  {author} {\bibinfo {author} {\bibfnamefont {T.}~\bibnamefont
  {Mizokawa}}\ and\ \bibinfo {author} {\bibfnamefont {A.}~\bibnamefont
  {Fujimori}},\ }\bibfield  {title} {\bibinfo {title} {Electronic structure and
  orbital ordering in perovskite-type 3$d$ transition-metal oxides studied by
  {Hartree-Fock} band-structure calculations},\ }\href
  {https://doi.org/10.1103/PhysRevB.54.5368} {\bibfield  {journal} {\bibinfo
  {journal} {Phys. Rev. B}\ }\textbf {\bibinfo {volume} {54}},\ \bibinfo
  {pages} {5368} (\bibinfo {year} {1996})}\BibitemShut {NoStop}%
\bibitem [{\citenamefont {Reuther}\ and\ \citenamefont
  {W\"olfle}(2010)}]{Reuther-2010}%
  \BibitemOpen
  \bibfield  {author} {\bibinfo {author} {\bibfnamefont {J.}~\bibnamefont
  {Reuther}}\ and\ \bibinfo {author} {\bibfnamefont {P.}~\bibnamefont
  {W\"olfle}},\ }\bibfield  {title} {\bibinfo {title}
  {{${J}_{1}\text{\ensuremath{-}}{J}_{2}$ frustrated two-dimensional Heisenberg
  model: Random phase approximation and functional renormalization group}},\
  }\href {https://doi.org/10.1103/PhysRevB.81.144410} {\bibfield  {journal}
  {\bibinfo  {journal} {Phys. Rev. B}\ }\textbf {\bibinfo {volume} {81}},\
  \bibinfo {pages} {144410} (\bibinfo {year} {2010})}\BibitemShut {NoStop}%
\bibitem [{\citenamefont {Baez}\ and\ \citenamefont
  {Reuther}(2017)}]{Baez2017}%
  \BibitemOpen
  \bibfield  {author} {\bibinfo {author} {\bibfnamefont {M.~L.}\ \bibnamefont
  {Baez}}\ and\ \bibinfo {author} {\bibfnamefont {J.}~\bibnamefont {Reuther}},\
  }\bibfield  {title} {\bibinfo {title} {Numerical treatment of spin systems
  with unrestricted spin length {$S$}: A functional renormalization group
  study},\ }\href {https://doi.org/10.1103/PhysRevB.96.045144} {\bibfield
  {journal} {\bibinfo  {journal} {Phys. Rev. B}\ }\textbf {\bibinfo {volume}
  {96}},\ \bibinfo {pages} {045144} (\bibinfo {year} {2017})}\BibitemShut
  {NoStop}%
\bibitem [{\citenamefont {Polchinski}(1984)}]{Polchinski1984}%
  \BibitemOpen
  \bibfield  {author} {\bibinfo {author} {\bibfnamefont {J.}~\bibnamefont
  {Polchinski}},\ }\bibfield  {title} {\bibinfo {title} {{Renormalization and
  effective Lagrangians}},\ }\href@noop {} {\bibfield  {journal} {\bibinfo
  {journal} {Nucl. Phys. B}\ }\textbf {\bibinfo {volume} {231}},\ \bibinfo
  {pages} {269} (\bibinfo {year} {1984})}\BibitemShut {NoStop}%
\bibitem [{\citenamefont {Wetterich}(1993)}]{Wetterich1993}%
  \BibitemOpen
  \bibfield  {author} {\bibinfo {author} {\bibfnamefont {C.}~\bibnamefont
  {Wetterich}},\ }\bibfield  {title} {\bibinfo {title} {{Exact evolution
  equation for the effective potential}},\ }\href@noop {} {\bibfield  {journal}
  {\bibinfo  {journal} {Phys. Lett. B}\ }\textbf {\bibinfo {volume} {301}},\
  \bibinfo {pages} {90} (\bibinfo {year} {1993})}\BibitemShut {NoStop}%
\end{thebibliography}%


%

\clearpage


\clearpage

\section{SUPPLEMENTARY MATERIAL}

\section{Methods}

\subsection{Sample preparation}

The powder of {\kni} was prepared by solid state reaction from a stoichiometric mixture of K$_2$SO$_4$ and NiSO$_4 \cdot 6$H$_2$O annealed at 450$^{\circ}$C for five days. The powder is quenched to room temperature and stored in a desiccator as {\kni} is mildly sensitive to moisture. High quality single crystals were obtained by sealing the powder in an evacuated quartz ampoule. Millimeter sized crystals are obtained by cooling the melt from 850$^{\circ}$C to 750$^{\circ}$C at a 1\,K/h rate.

\subsection{Single-crystal x-ray diffraction}

A small single crystal of {\kni} has been glued onto the tip of a glass needle and cooled down to 100 K with a flow of cold nitrogen gas. Data has been collected on a Rigaku SuperNOVA diffractometer using Mo/Cu Duo source with Atlas CCD.

\subsection{Magnetization and magnetic susceptibility}

Magnetization $M$ and magnetic susceptibility $\chi_\mathrm{DC} = M/B$ of powder and single crystal samples were measured using a commercial superconducting quantum interference device magnetometer MPMS-5T (Quantum Design).

\subsection{Heat capacity}

Heat capacity measurements above 2\,K were performed on powder and single crystal samples using a commercial PPMS (Quantum Design). Below 2\,K, a home-made setup using a dilution refrigerator has been used to measure single crystal sample. In both cases a short (1-3 \%) heat pulse method has been utilized.

\subsection{Muon spin relaxation ($\mu$SR)}

$\mu$SR experiments were performed on powder samples at MUSR, ISIS (UK) and LTF/GPS, PSI (Switzerland) beamlines using the spin-polarized positive muons ($\mu^+$).

\subsection{Neutron diffraction}

Neutron diffraction on powder was performed on the time-of-flight diffractometer WISH, ISIS (UK). For temperatures below 1\,K, a copper can was attached to a dilution refrigerator and filled with 15 g of powder. Above 1\,K, a vanadium can was used with 15 g of powder in a helium-flow environment.

\subsection{Spin-polarized neutron diffraction and inelastic neutron scattering}

Both spin-polarized neutron diffraction and non-polarized time-of-flight (TOF) inelastic neutron scattering measurements were carried out at the polarized spectrometer DNS at the Heinz Maier-Leibnitz Zentrum (MLZ), Garching, Germany. Approximately 2 g of powder were enclosed in an annular cylinder sample holder made with oxygen-free copper and sealed in a He atmosphere. Measurements were taken in a $^3$He insert installed in a top-loading CCR cryostat. A neutron wavelength at $\lambda=4.2$ {\AA} was chosen for both measurements. The magnetic scattering cross-section was obtained via the XYZ polarization analysis method, for which the standard procedures such as flipping-ratio correction and normalisation of detector efficiency have been applied. The TOF inelastic neutron scattering data were taken with a disc chopper running at 250 Hz, which yields an energy resolution at $\sim$0.25 meV at 4.2 {\AA}. The runs for both vanadium and empty copper sample can were undertaken under the same TOF condition. The powder-average inelastic scattering profiles were obtained via Mantid-based data reduction routines.

\subsection{Density functional theory}

We study {\kni} using density functional theory (DFT) calculations based on the full potential local orbital (FPLO) basis set~\cite{Koepernik1999} combined with the generalized gradient approximation (GGA) to the exchange correlation functional~\cite{Perdew1996} and with a GGA+$U$ correction for the strongly correlated Ni$^{2+}$ $3d$ orbitals~\cite{Liechtenstein1995}. We employ the energy mapping technique~\cite{Jeschke2015,Jeschke2019} to extract the Heisenberg exchange interactions up to a Ni-Ni distance of 8.6\,{\AA} from 20 GGA+$U$ total energies of selected spin configurations in a $\sqrt{2}\times\sqrt{2}\times 1$ supercell. We fix the Hund's rule coupling at $J_{\rm H}=0.88$\,eV following Ref.~\cite{Mizokawa1996}.

\subsection{PFFRG}

The model Hamiltonian for K$_2$Ni$_2$(SO$_4$)$_3$ with the Heisenberg exchange interactions obtained from DFT is further studied within the pseudofermion functional renormalization group (PFFRG) method.~\cite{Reuther-2010} This approach is based on a fermionic rewriting of the spin operators, where a spin-1 is represented by two coupled spin-1/2 degrees of freedom.~\cite{Baez2017} The resulting fermionic theory is then treated with many-body Feynman diagram approaches. Particularly, via the introduction of an infrared frequency cutoff, the fermionic vertex functions are subject to a renormalization group flow as described within the standard functional renormalization group (FRG) scheme.~\cite{Polchinski1984,Wetterich1993} We solve the corresponding differential equations in real space on a one-loop level, by taking into account spin-spin correlations up to a distance of twice a lattice vector of the underlying cubic lattice and approximate the frequency dependence of the vertex functions by 64 discrete mesh points. The central outcome is the zero-frequency, momentum-resolved real part of the magnetic susceptibility $\chi'(Q)$ which is obtained from the fermionic two-particle vertex. Using Kramers-Kronig relations, $\chi'(Q)$ is related to the dynamical spin structure factor $S(Q,\omega)$ via
\begin{equation}
\chi'(Q)\propto\int d\omega S(Q,\omega)/\omega\;,
\end{equation}
indicating that $\chi'(Q)$ primarily represents the low-energy contribution of $S(Q,\omega)$. Most importantly, $\chi'(Q)$ takes into account quantum fluctuations well beyond mean field and is, hence, well suited to simulate the fluctuating moments of K$_2$Ni$_2$(SO$_4$)$_3$. Furthermore, possible instability signatures during the renormalization group flow allow one to detect static magnetic long-range order.

\subsection{Classical Monte Carlo}

Monte Carlo simulations are performed for classical Heisenberg spins on the bi-trillium lattice with periodic boundary conditions for a system of $8L^{3}$ spins. We employ the single-flip metropolis update with 5 over-relaxation steps added after every sweep of the lattice, and $10^{4}$ Monte Carlo sweeps are used for thermalization. This is followed by $10^{5}$ Monte Carlo sweeps during which measurements are performed every $10$ Monte Carlo sweeps. The calculations for magnetization as a function of applied field shown in Fig.~2(b) of the main text
are performed for a lattice size of $L=8$ ($4096$ spins).

\section{x-ray diffraction}

\begin{figure*}[h]
    \centering
    \includegraphics[width=0.9\columnwidth]{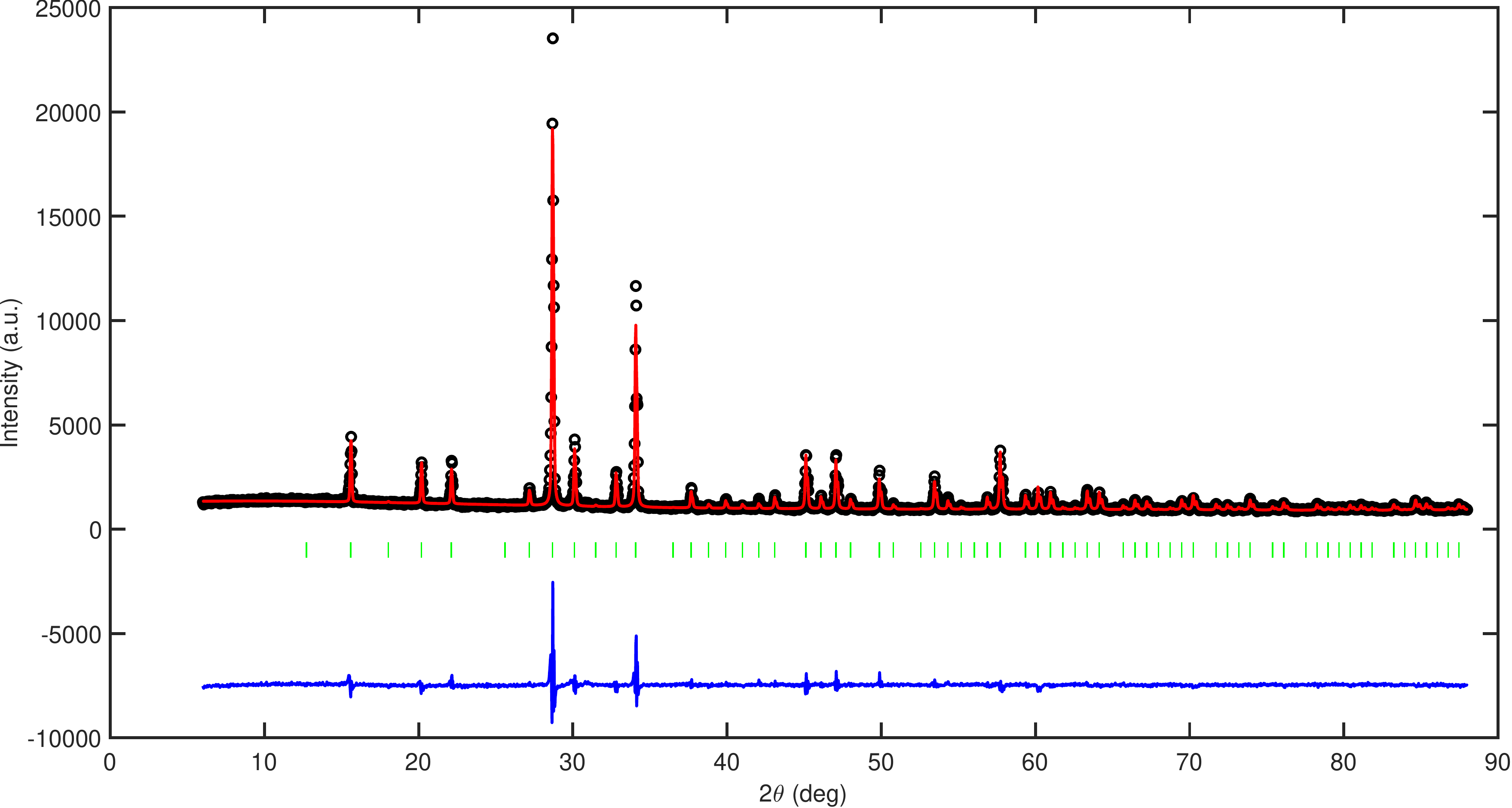}
    \caption{Powder diffraction of {\kni} (black circles) with the result of a Rietveld refinement (red line). The difference between the measured intensities and the fit is given with the blue line. Peak positions are marked with green vertical lines.}
    \label{fig:powder}
\end{figure*}

Powder x-ray diffraction of {\kni} at room temperature is presented in Fig.~\ref{fig:powder}. The agreement is very good ($R_{wp} = 6.2$ \%), with no visible traces of impurities.

\begin{figure}[h]
    \centering
    \includegraphics[width=0.9\columnwidth]{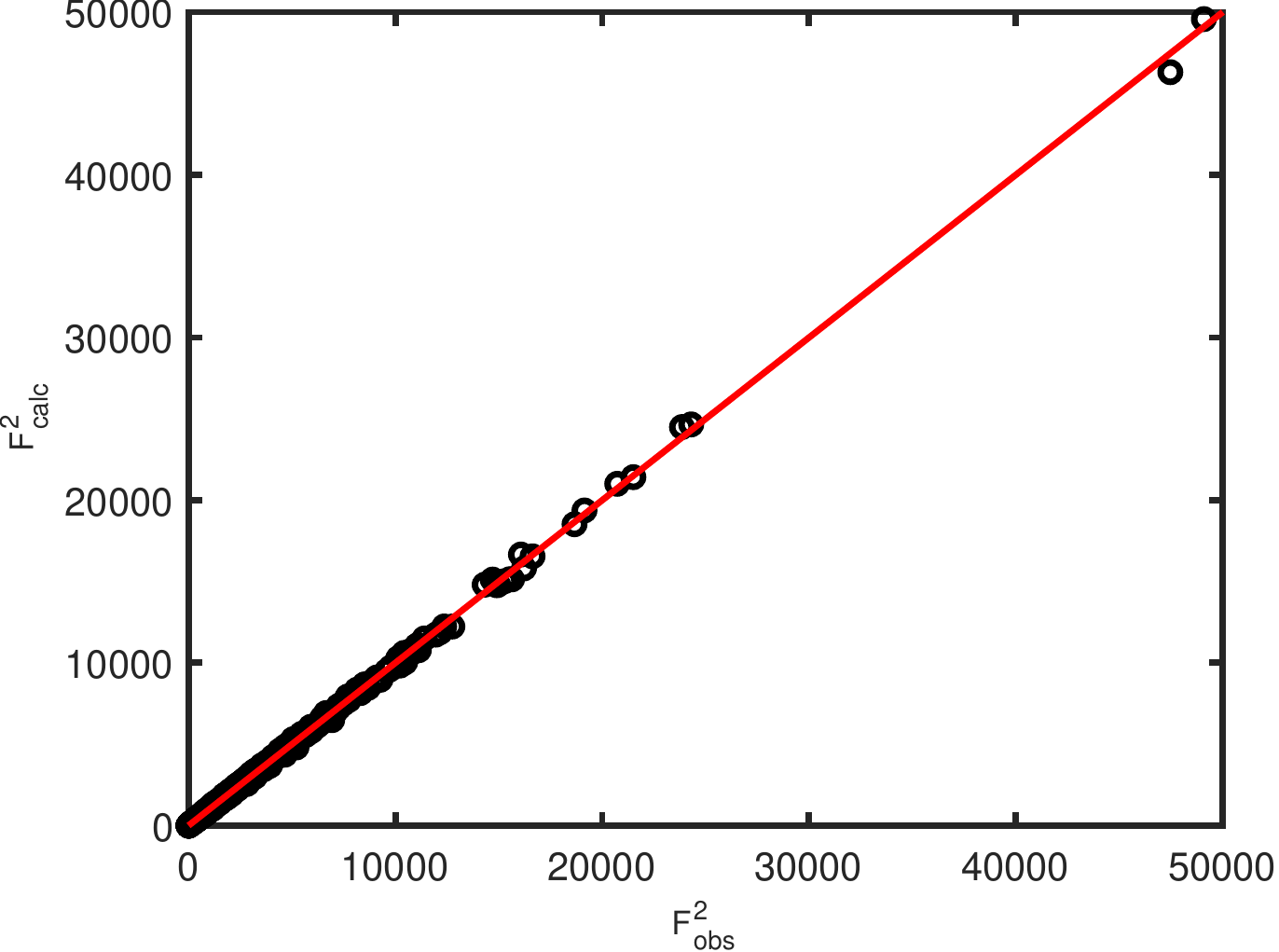}
    \caption{Experimental structure factors are plotted against the calculated structure factors obtained by single crystal structure refinement of {\kni}.}
    \label{fig:SCxray}
\end{figure}

The experimental versus calculated structure factors for a single crystal of {\kni} is shown in Fig.~\ref{fig:SCxray}. The tight distribution of the data around the red line $F^2_{obs} = F^2_{calc}$ indicates the high quality of the refinement.
Additional refinement parameters are given in the supplementary Table~\ref{table:SCquality}. The agreement factors and the goodness-of-fit value confirm the high accuracy of the {\kni} structure description. Atomic positions as well as distances and angles are listed in supplementary Table~\ref{table:coordinates} and~\ref{table:distances}, respectively.

\begin{table*}[h]
\begin{tabular}{|l|l|}
\hline
Temperature                                    & 100.01(10) K                                                                                 \\ \hline
Crystal system, space group                    & Cubic,  P2(1)3                                                                               \\ \hline
a = b = c                                      & 9.81866(12) A                                                                                \\ \hline
Volume                                         & 946.58(4) A3                                                               \\ \hline
Z, Calculated density                          & 3,  3.395 g/cm$^3$                                                            \\ \hline
Absorption coefficient                         & 5.589 mm$^{-1}$                                                                 \\ \hline
F(000)                                         & 952                                                                                          \\ \hline
Theta range for data collection                & 3.594 to 30.444 deg                                                                         \\ \hline
Limiting indices                               & -5$\leq$h$\leq$14, -9$\leq$k$\leq$14, -14$\leq$l$\leq$13 \\ \hline
Reflections collected / unique                 & 4119 / 965 R(int) = 0.0274                                                             \\ \hline
Completeness to $\theta$ = 25.242                 & 98.8 \%                                                                                      \\ \hline
Data / restraints / parameters                 & 965 / 0 / 59                                                                                 \\ \hline
Goodness-of-fit on F$^2$        & 1.046                                                                                        \\ \hline
Final R indices I$>$2sigma(I) & R1 = 0.0143, wR2 = 0.0330                                                                    \\ \hline
R indices (all data)                           & R1 = 0.0146, wR2 = 0.0332                                                                    \\ \hline
Absolute structure parameter                   & -0.034(9)                                                                                    \\ \hline
Extinction coefficient                         & 0.0145(8)                                                                                    \\ \hline
Largest diff. peak and hole                    & 0.238 and -0.245 e.A$^{-3}$                                                    \\ \hline

\end{tabular}
\caption{Single crystal refinement parameters of the {\kni} structure at 100 K}
\label{table:SCquality}
\end{table*}

\begin{table}[]
\begin{tabular}{|l|l|l|l|l|l|}
\hline
      & x       & y        & z       & U(eq) & Site \\ \hline
Ni(1) & 1645(1) & 1645(1)  & 1645(1) & 5(1)  & 4a   \\ \hline
Ni(2) & 5945(1) & 945(1)   & 4055(1) & 5(1)  & 4a   \\ \hline
K(1)  & 1854(1) & -1854(1) & 3146(1) & 10(1) & 4a   \\ \hline
K(2)  & 4507(1) & 4507(1)  & 4507(1) & 10(1) & 4a   \\ \hline
S(1)  & 2826(1) & 1233(1)  & 4806(1) & 5(1)  & 12b  \\ \hline
O(1)  & 2550(2) & 952(2)   & 3371(2) & 12(1) & 12b  \\ \hline
O(2)  & 2581(2) & -28(2)   & 5572(2) & 15(1) & 12b  \\ \hline
O(3)  & 4246(2) & 1699(2)  & 4987(2) & 10(1) & 12b  \\ \hline
O(4)  & 1907(2) & 2262(2)  & 5371(2) & 13(1) & 12b  \\ \hline
\end{tabular}
\caption{Fractional atomic coordinates ($\times 10^4$) and equivalent isotropic displacement parameters (\AA$^2 \times 10^3$) for {\kni} at 100 K. $U(eq)$ is defined as one third of the trace of the orthogonalized $U_{ij}$ tensor. }
\label{table:coordinates}
\end{table}

\begin{table}[]
\begin{tabular}{|l|l|}
\hline
Ni(1)-K(1)      & 3.7435(3)  \\ \hline
Ni(1)-O(1)      & 2.0315(18) \\ \hline
Ni(2)-K(2)      & 3.7977(4)  \\ \hline
Ni(2)-O(3)      & 2.0417(18) \\ \hline
K(1)-O(1)       & 2.847(2)   \\ \hline
K(1)-O(2)       & 3.065(2)   \\ \hline
K(2)-O(3)       & 2.8094(19) \\ \hline
S(1)-O(1)       & 1.4613(18) \\ \hline
S(1)-O(2)       & 1.469(2)   \\ \hline
S(1)-O(3)       & 1.4782(18) \\ \hline
S(1)-O(4)       & 1.4636(18) \\ \hline
\end{tabular}
\caption{Bond lengths (in \AA) for {\kni} at 100 K.}
\label{table:distances}
\end{table}

\begin{table}[]
\begin{tabular}{|l|l|}
\hline
O(1)-Ni(1)-K(1) & 48.72(6)   \\ \hline
O(3)-Ni(2)-K(2) & 46.38(5)   \\ \hline
O(1)-K(1)-O(2)  & 47.00(5)   \\ \hline
K(1)-S(1)-K(2)  & 146.85(2)  \\ \hline
O(1)-S(1)-K(1)  & 49.47(8)   \\ \hline
O(1)-S(1)-K(2)  & 100.00(8)  \\ \hline
O(1)-S(1)-O(2)  & 107.72(12) \\ \hline
O(1)-S(1)-O(3)  & 110.41(11) \\ \hline
O(1)-S(1)-O(4)  & 112.40(11) \\ \hline
O(2)-S(1)-K(1)  & 58.29(9)   \\ \hline
O(2)-S(1)-K(2)  & 149.72(9)  \\ \hline
O(2)-S(1)-O(3)  & 110.71(11) \\ \hline
O(3)-S(1)-K(1)  & 124.71(7)  \\ \hline
O(3)-S(1)-K(2)  & 46.05(7)   \\ \hline
O(4)-S(1)-K(1)  & 126.40(8)  \\ \hline
O(4)-S(1)-K(2)  & 72.52(8)   \\ \hline
O(4)-S(1)-O(2)  & 106.69(12) \\ \hline
O(4)-S(1)-O(3)  & 108.85(11) \\ \hline
Ni(1)-O(1)-K(1) & 98.86(7)   \\ \hline
S(1)-O(1)-Ni(1) & 145.35(12) \\ \hline
S(1)-O(1)-K(1)  & 107.57(10) \\ \hline
S(1)-O(2)-K(1)  & 97.65(10)  \\ \hline
Ni(2)-O(3)-K(2) & 101.88(7)  \\ \hline
S(1)-O(3)-Ni(2) & 127.24(10) \\ \hline
S(1)-O(3)-K(2)  & 111.68(9)  \\ \hline
\end{tabular}
\caption{Angles (in degrees) for {\kni} at 100 K.}
\label{table:distances}
\end{table}

\clearpage
\section{Magnetization}

Low temperature view of the temperature dependence of the inverse of magnetic susceptibility $\chi_{dc} = M/B$. A weak deviation from the Curie-Weiss law starts below 50 K but it is significantly visible only below 20 K.

\clearpage
\section{Specific heat}

As shown in Supplementary Figure~\ref{fig:kinks}, at higher temperatures both {\kni} and the non-magnetic analog {\kmg} show kinks in their specific heat. These are probably related to the freezing of SO$_4$ groups without a noticeable symmetry lowering from the cubic space group.

The phonon contribution below 2 K has been estimated by employing a polynomial $BT^3 + CT^5$ with $B = 1.33(3) \cdot 10^{-3}$ J/mol K$^4$ and $C = 2.1(8) \cdot 10^{-6}$ J/mol K$^6$ that best matches the measured specific heat of {\kmg} at low temperatures, as seen in Supplementary Figure~\ref{fig:lowTextension}.

\clearpage
Determination of specific heat involved in the second order phase transition at $T^*$ is displayed in Supplementary Figure~\ref{fig:Tstar}. The red curved dashed line is the measurement at 5 T where no anomaly is present, adjusted to match the zero-field data at 0.8 K and 1.5 K. The exact position of the background line does not change significantly the extracted value of $\sim 1$ \% of the total $R ln(3)$ entropy of spin-1 system.

Magnetic field evolution of $T^*$ and $T^{**}$ is presented in Supplementary Figure~\ref{fig:Bdependence}. For $B = 0.75$ T a small shoulder appears around 0.5 K, possibly indicating another phase. Given that this is seen for a very narrow magnetic field range, it could also reflect an experimental artifact.

\clearpage
Comparison of a power-law behavior $C_P \sim T^n$ for $n = 2$ and $n = 3$ with a gaped behavior $C_P \sim exp(-\Delta/T)$ is presented in Supplementary Figure~\ref{fig:slopes}. Zero field data show a somewhat varying slope, possibly influenced by the presence of the static component. $\Delta = 0.5$ K corresponds to the value of a spin-triplet gap for a static dimer on the $J_4$ bond.

Comparison between results of specific heat obtained on single crystal and on powder samples is shown in Supplementary Figure~\ref{fig:SCpowder}.

\clearpage
\section{Neutron scattering}

In Fig.~\ref{fig:DNS} we show the temperature dependence of the scattering profile of polarized neutrons. At 17 K the period of oscillations is still visible, although with a smaller amplitude. At 80 K the profile is practically featureless. The spikes occur at positions of nuclear Bragg peaks and are related to imperfect subtraction of large numbers in spin-flip and non-spin-flip channels.

Fig.~\ref{fig:LeBail} displays the diffraction data at 90 mK together with a LeBail fit using three propagation vectors $Q_1 = (\frac{1}{3},0,0)$, $Q_2 = (\frac{1}{3},\frac{1}{3},0)$ and $Q_3 = (\frac{1}{3},\frac{1}{3},\frac{1}{3})$. For $Q > 1$ \AA ~the satellites appear as shoulders of strong nuclear Bragg peaks which together with a diminishing form factor makes them very hard to distinguish.

\clearpage
\section{Muon spin relaxation}

Taking into account that the whole data set is measured with two different experimental setups (a dilution refrigerator and a variable temperature insert, implying different backgrounds and different initial asymmetries), the temperature evolution of relaxation rates and exponents are presented in two segments. The low temperature segment, from 20 mK up to 4 K, is modeled using Eq.(2) from the main text, while the high temperature segment, from 100 K down to 3 K, is modeled using Eq.(1). In the region around 3 K both approaches can be used so if the low temperature segment is modeled using Eq.(1), the extracted relaxation rates overlap, as shown in Fig.~\ref{fig:muons2equations} with green and blue diamonds.

It is rather simple to understand why two approaches work equally well. In this overlapping region the exponent $\beta$ acquires values close to 1, rendering two contributions in Eq.(2) identical and effectively becoming Eq.(1). For two approaches to smoothly transform from one to the other it would be necessary to allow for the fraction $f$ and $\beta_1$ to be freely varied or that a microscopic model is developed which could meaningfully constrain other parameters.

Fig.~\ref{fig:muons} shows longitudinal-field $\mu$SR relaxation at 1.7 K, well within the correlated region. The system remains dynamic with fields up to 0.78 T.

\clearpage
\section{Additional PFFRG information}

In PFFRG the magnetic susceptibility depends on the renormalization group parameter $\Lambda$ which is implemented as a sharp infrared frequency cutoff. Despite the artificial nature of $\Lambda$ it shows various similarities with the temperature $T$, particularly, kinks or cusps in the $\Lambda$-dependence of the susceptibility signal the onset of magnetic long-range order. Most importantly, the identification of either magnetic long-range order or a magnetically disordered phase does not rely on any prior assumption on the system's ground state. To illustrate the identification of magnetic order, Fig.~\ref{fig:PFFRGlambda} shows the maximal susceptibility in momentum space as a function of $\Lambda$ for various different systems. The orange curve corresponds to a spin-1 Heisenberg model on the lattice network of K$_2$Ni$_2$(SO$_4$)$_3$ but with $J_4>0$ interactions only. The antiferromagnetic N\'eel order in this system manifests in a strong peak. On the other hand, the green curve is a
typical example for a smooth non-magnetic renormalization group flow as given for the spin-$1/2$ nearest neighbor antiferromagnetic Heisenberg model on the pyrochlore lattice. The PFFRG data for K$_2$Ni$_2$(SO$_4$)$_3$ is presented by the blue curve and shows an
intermediate behavior: A small kink at $\Lambda\approx0.45$ is observed which, however, does not develop into a pronounced peak (note that small oscillations below $\Lambda\approx0.45$ are typically artifacts of the discretization of continuous frequency variables within our numerics). This indicates that our PFFRG results are in accord with a small ordered moment in the absence of an external magnetic field.

\begin{figure}[h]
    \centering
    \includegraphics[width=0.9\columnwidth]{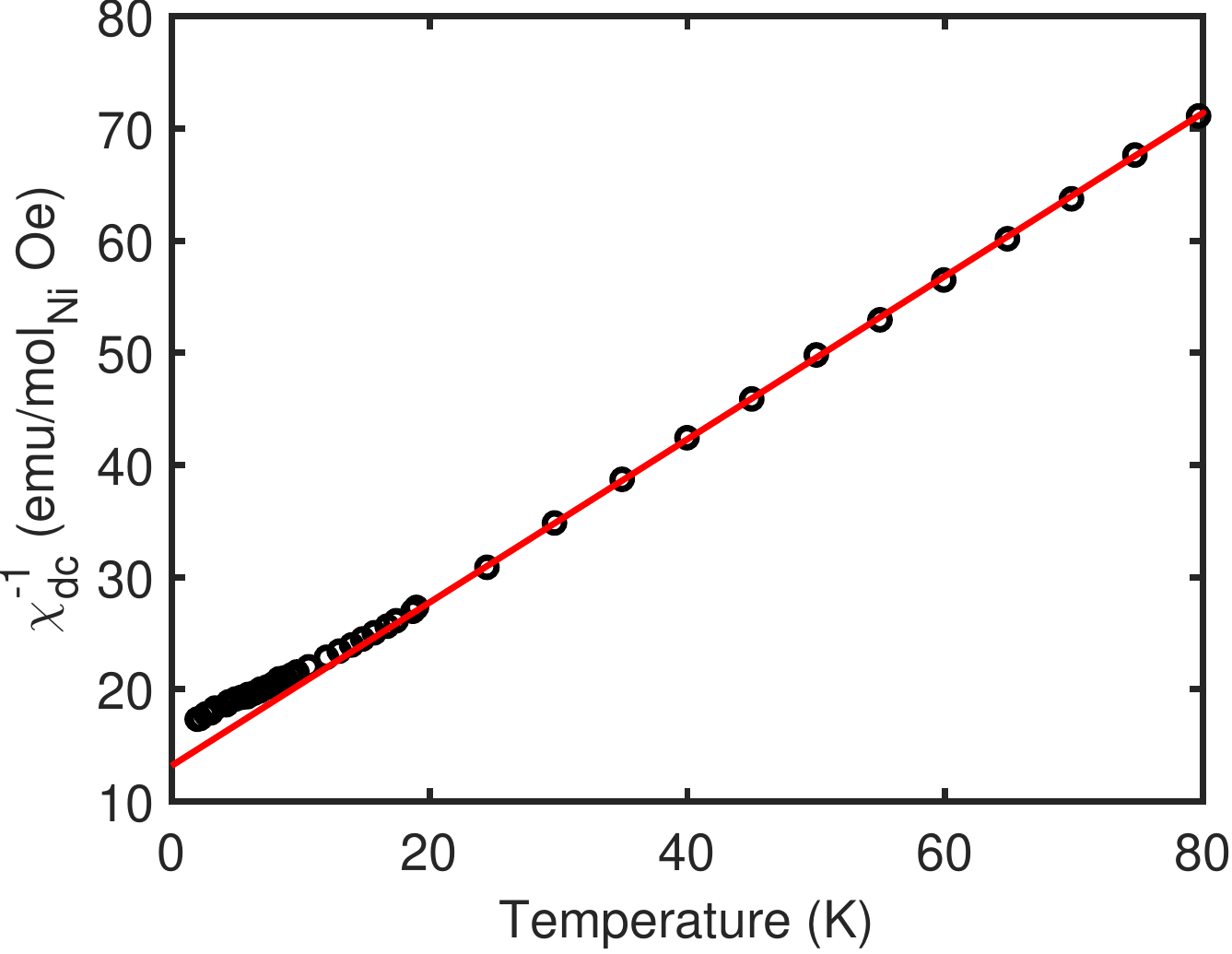}
    \caption{Deviation from the Curie-Weiss law.}
    \label{fig:CWzoom}
\end{figure}
\begin{figure}[h]
    \centering
    \includegraphics[width=0.9\columnwidth]{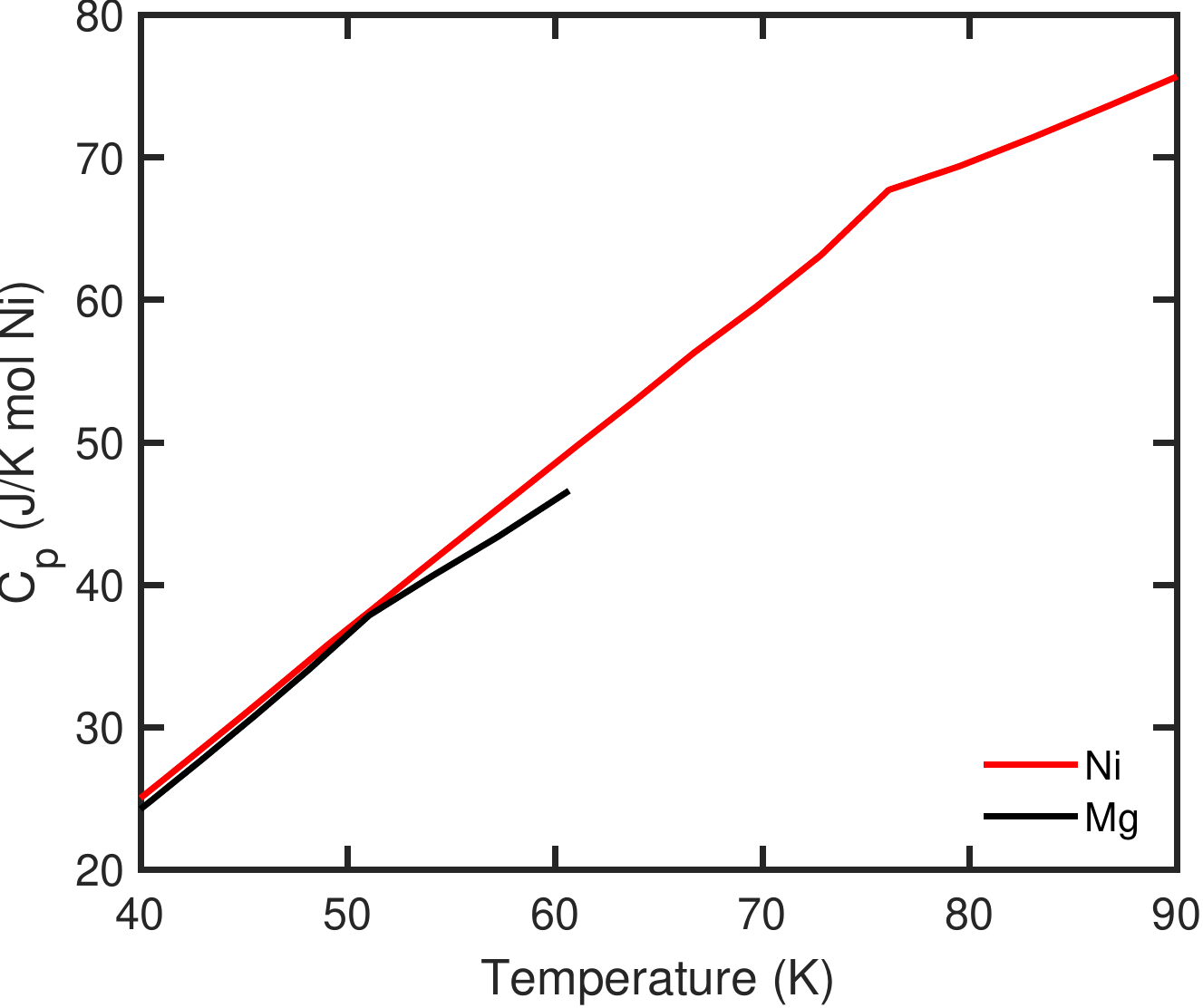}
    \caption{Kinks in the specific heat of {\kni} and {\kmg}.}
    \label{fig:kinks}
\end{figure}
\begin{figure}[h]
    \centering
    \includegraphics[width=0.9\columnwidth]{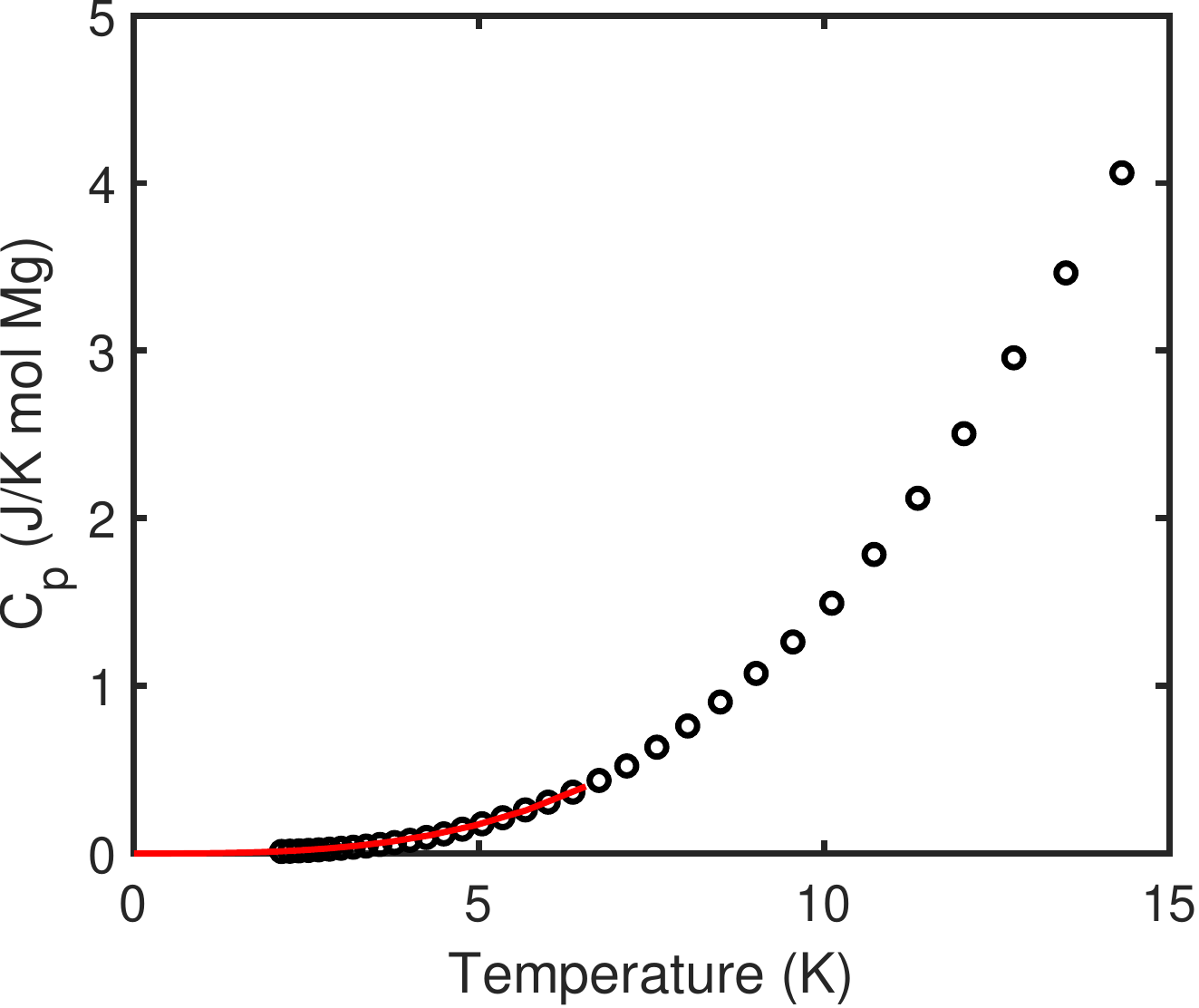}
    \caption{Specific heat of {\kmg} with a low temperature extension based on a polynomial $BT^3 + CT^5$.}
    \label{fig:lowTextension}
\end{figure}
\begin{figure}[h]
    \centering
    \includegraphics[width=0.8\columnwidth]{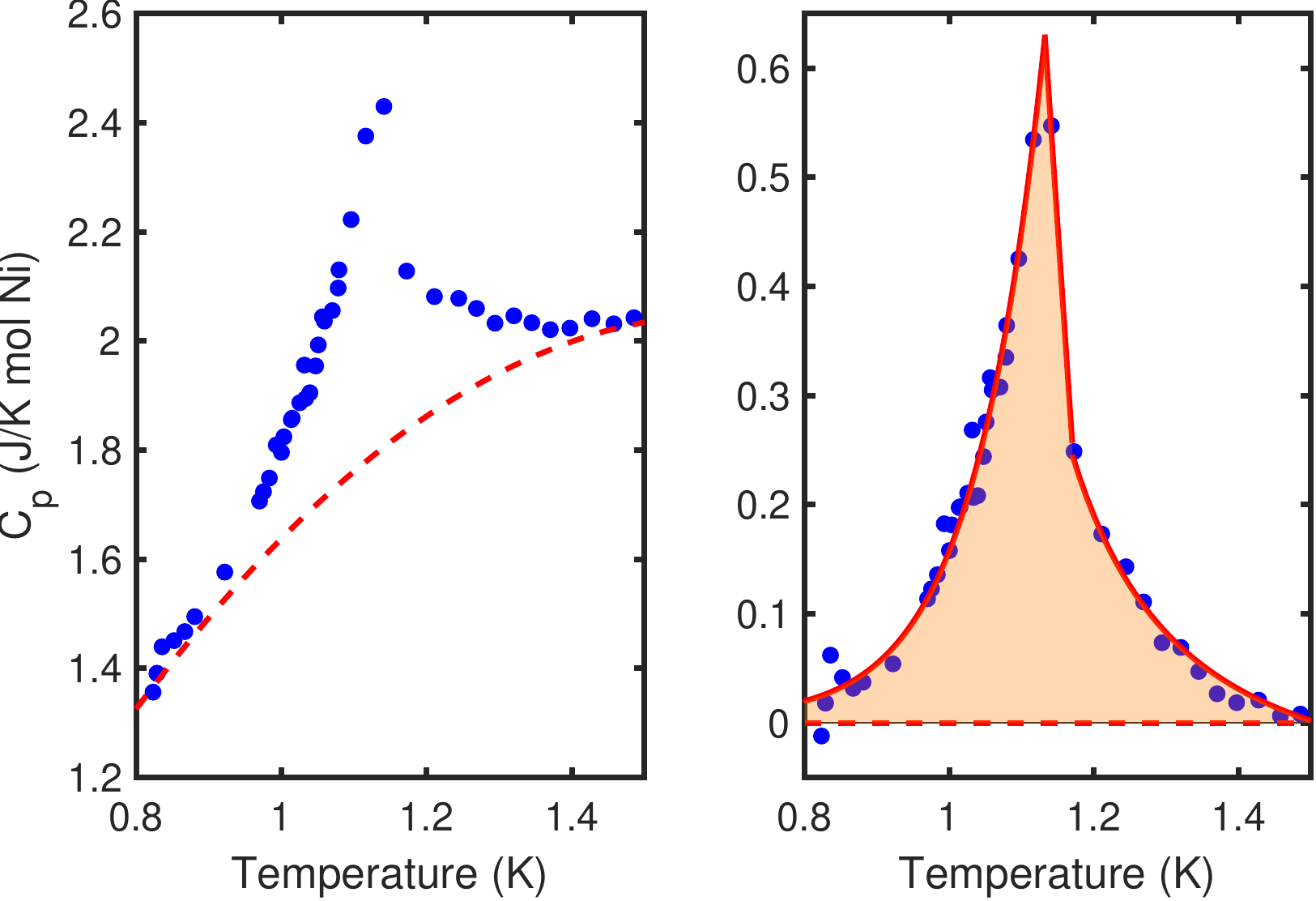}
    \caption{Second order transition at $T^*$. The shaded area carries an entropy of $\sim 1$ \% of the total $Rln3$.}
    \label{fig:Tstar}
\end{figure}
\begin{figure}[h]
    \centering
    \includegraphics[width=0.9\columnwidth]{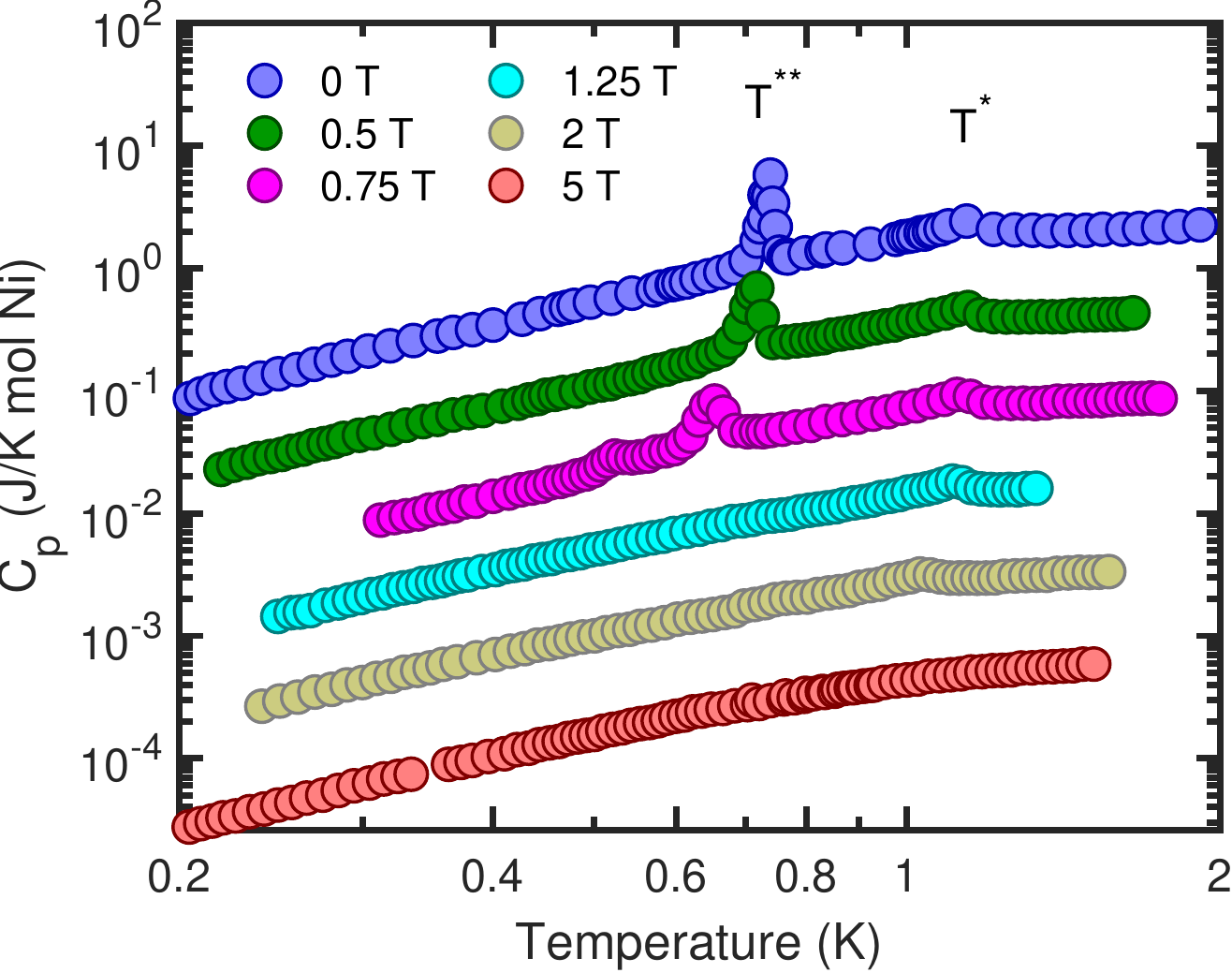}
    \caption{Magnetic field dependence of specific heat below 2\,K. Individual curves are shifted vertically for clarity.}
    \label{fig:Bdependence}
\end{figure}
\begin{figure}[h]
    \centering
    \includegraphics[width=0.9\columnwidth]{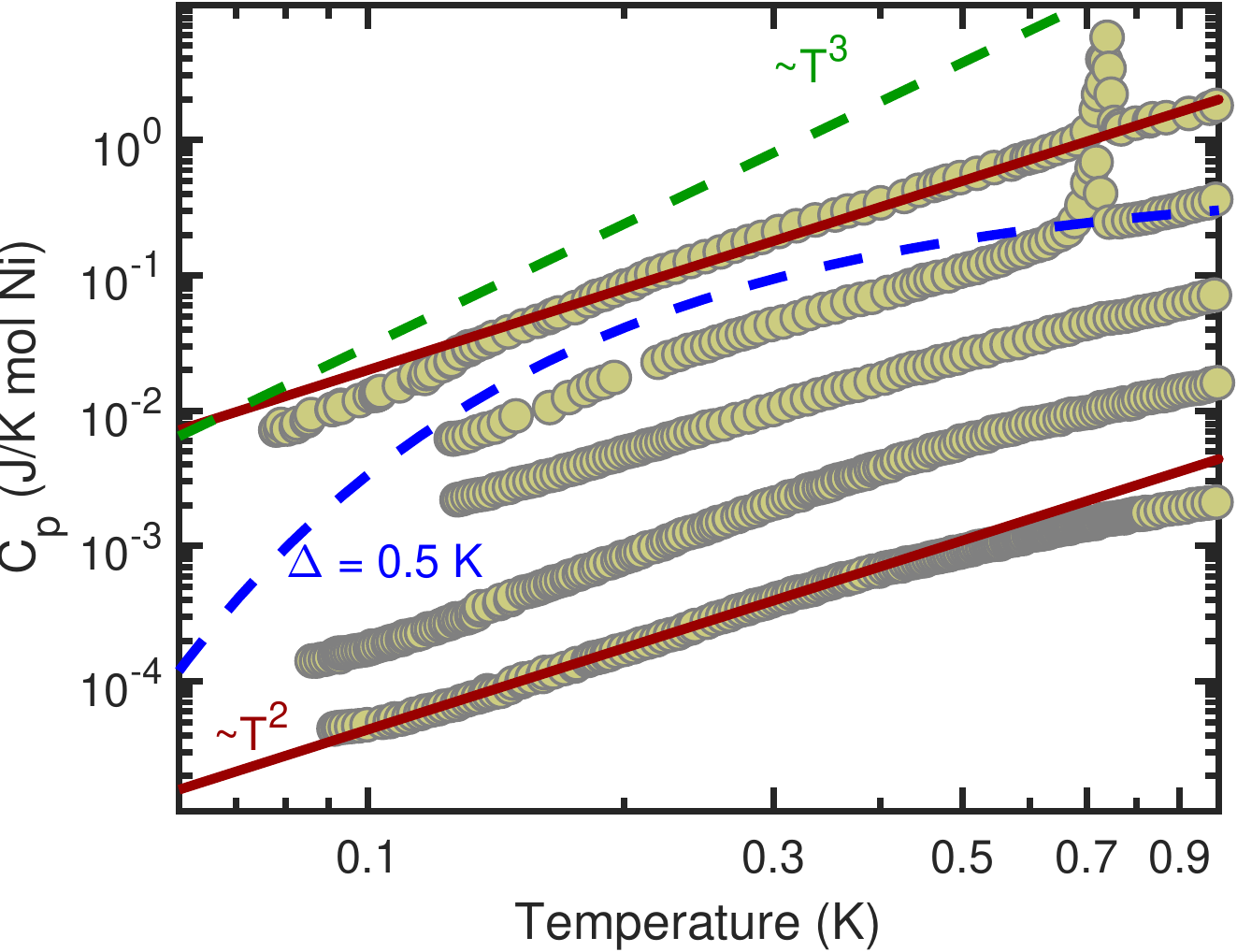}
    \caption{Low temperature specific heat for magnetic field values $B = 0, 0.5, 1.5, 7$ and $14$ T (top to bottom). Individual curves are shifted vertically for clarity.}
    \label{fig:slopes}
\end{figure}
\begin{figure}[h]
    \centering
    \includegraphics[width=0.9\columnwidth]{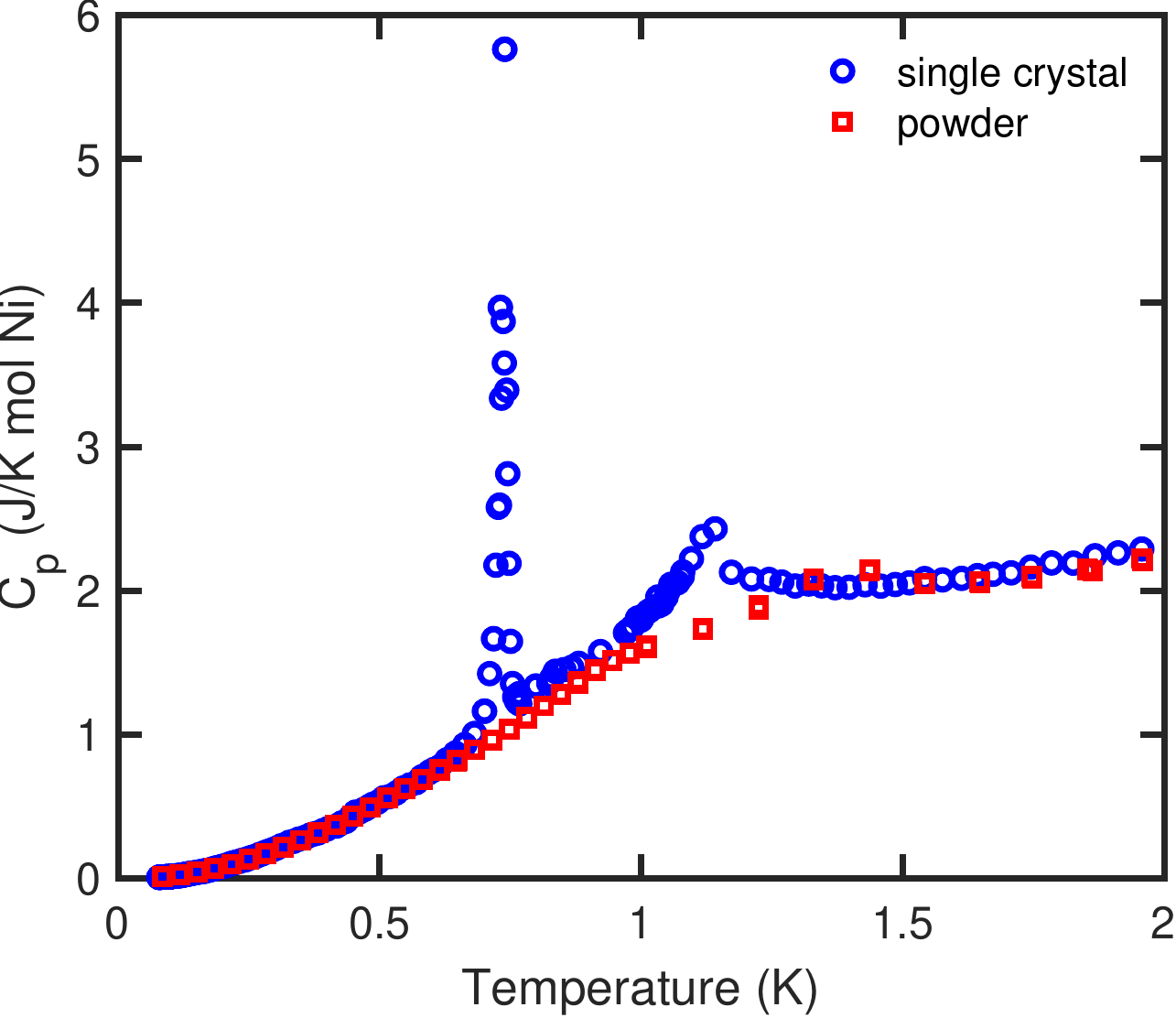}
    \caption{Comparison between single crystal and powder measurements.}
    \label{fig:SCpowder}
\end{figure}
\begin{figure}[h]
    \centering
    \includegraphics[width=0.9\columnwidth]{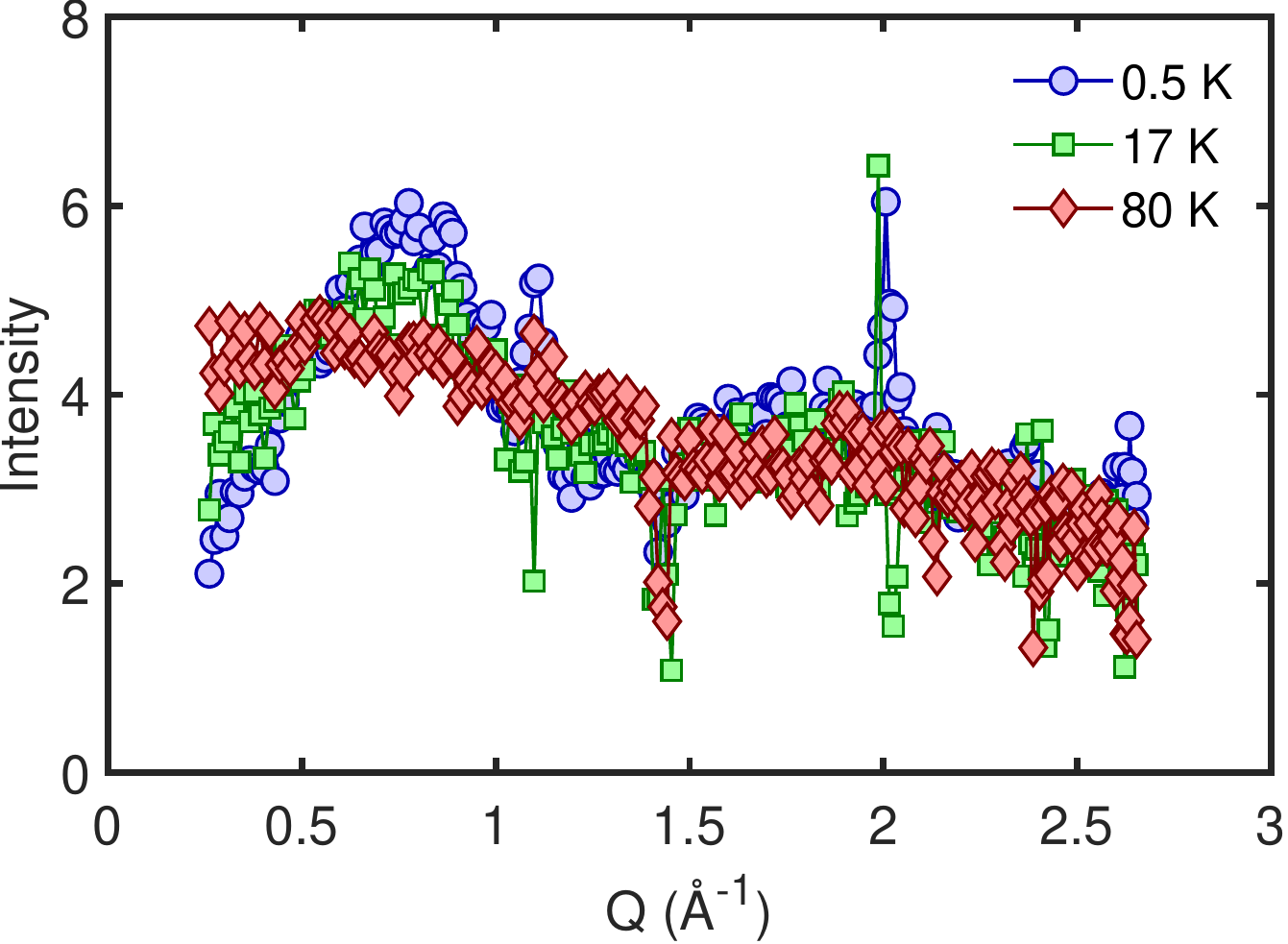}
    \caption{Temperature dependence of spin-polarized neutron diffraction.}
    \label{fig:DNS}
\end{figure}
\begin{figure}[h]
    \centering
    \includegraphics[width=0.9\columnwidth]{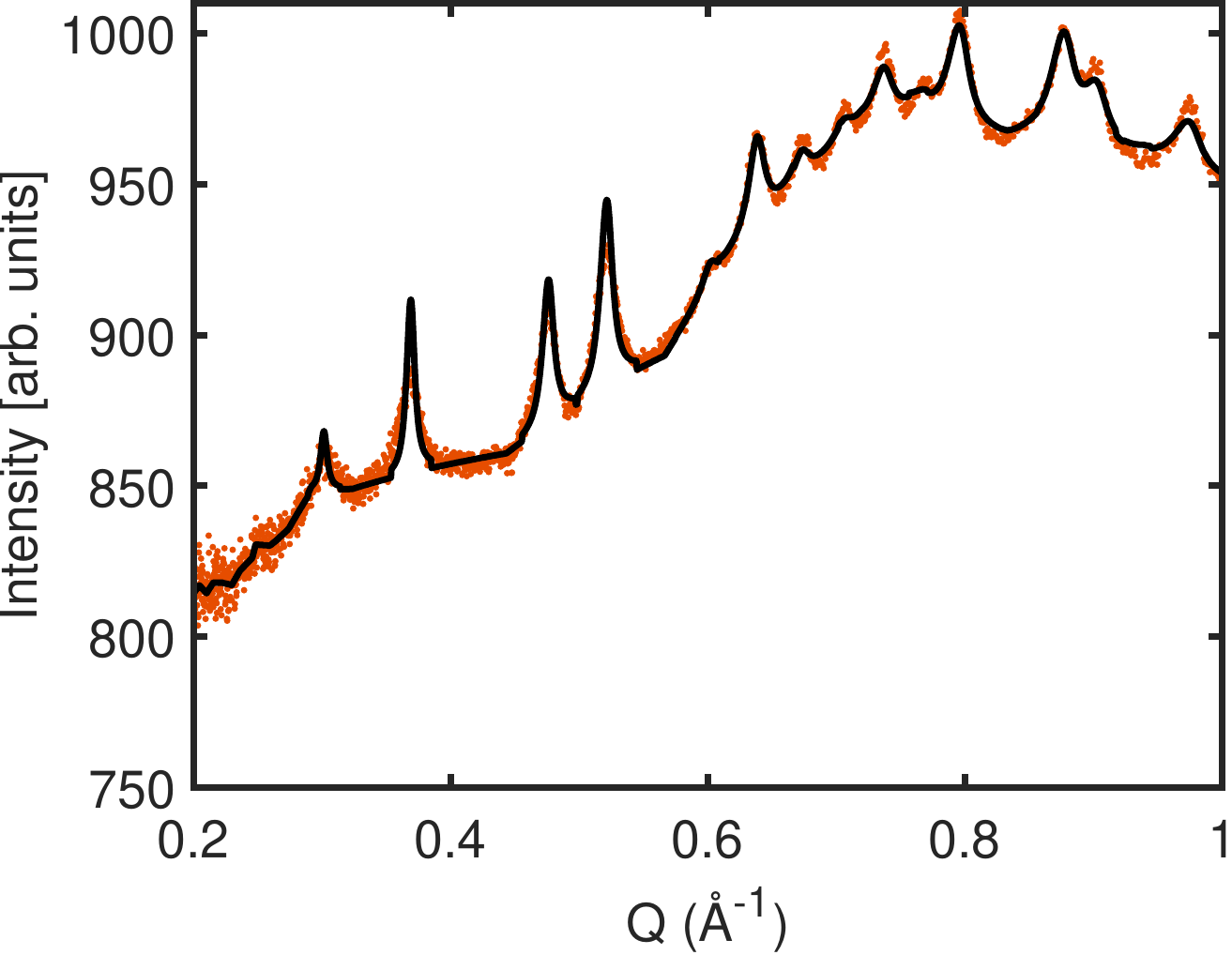}
    \caption{Low-$Q$ diffraction profile of {\kni} at 90 mK (points) with a Lebail fit (line).}
    \label{fig:LeBail}
\end{figure}
\begin{figure}[h]
    \centering
    \includegraphics[width=0.5\columnwidth]{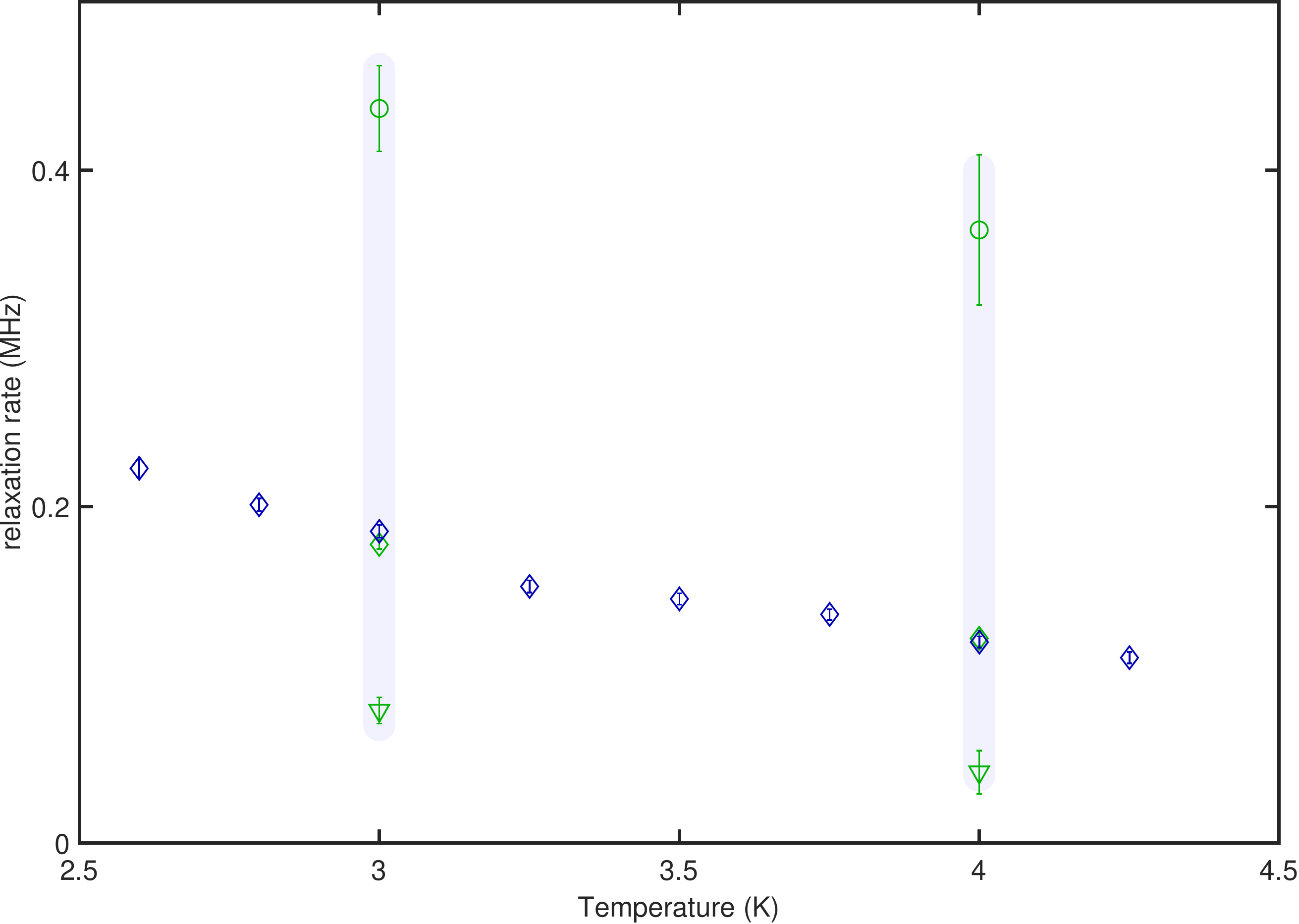}
    \caption{Temperature region where two segments are equally well described with both equations (see main text).}
    \label{fig:muons2equations}
\end{figure}
\begin{figure}[h]
    \centering
    \includegraphics[width=0.9\columnwidth]{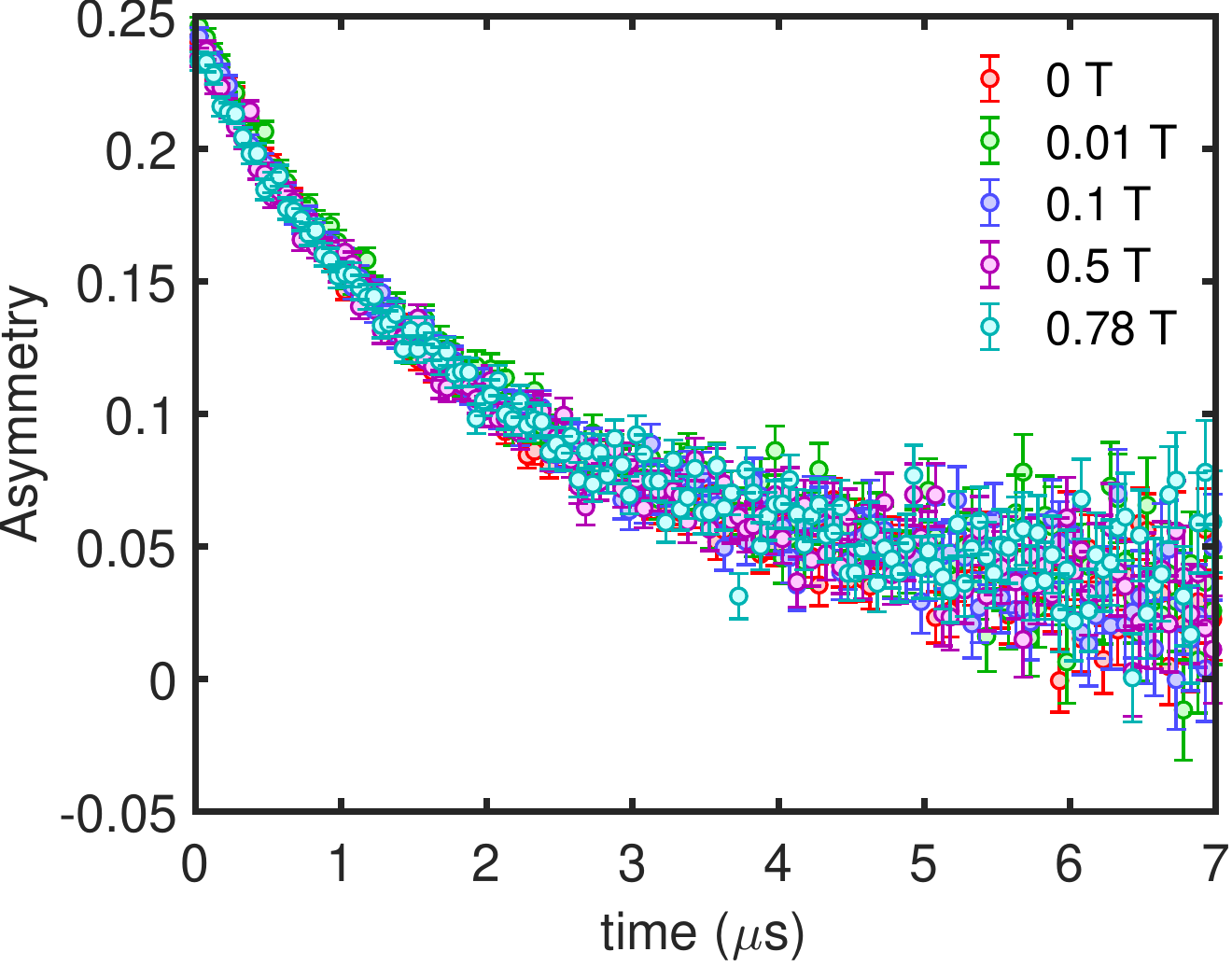}
    \caption{Longitudinal-field $\mu$SR relaxation at 1.7 K.}
    \label{fig:muons}
\end{figure}
\begin{figure}[h]
    \centering
    \includegraphics[width=0.9\columnwidth]{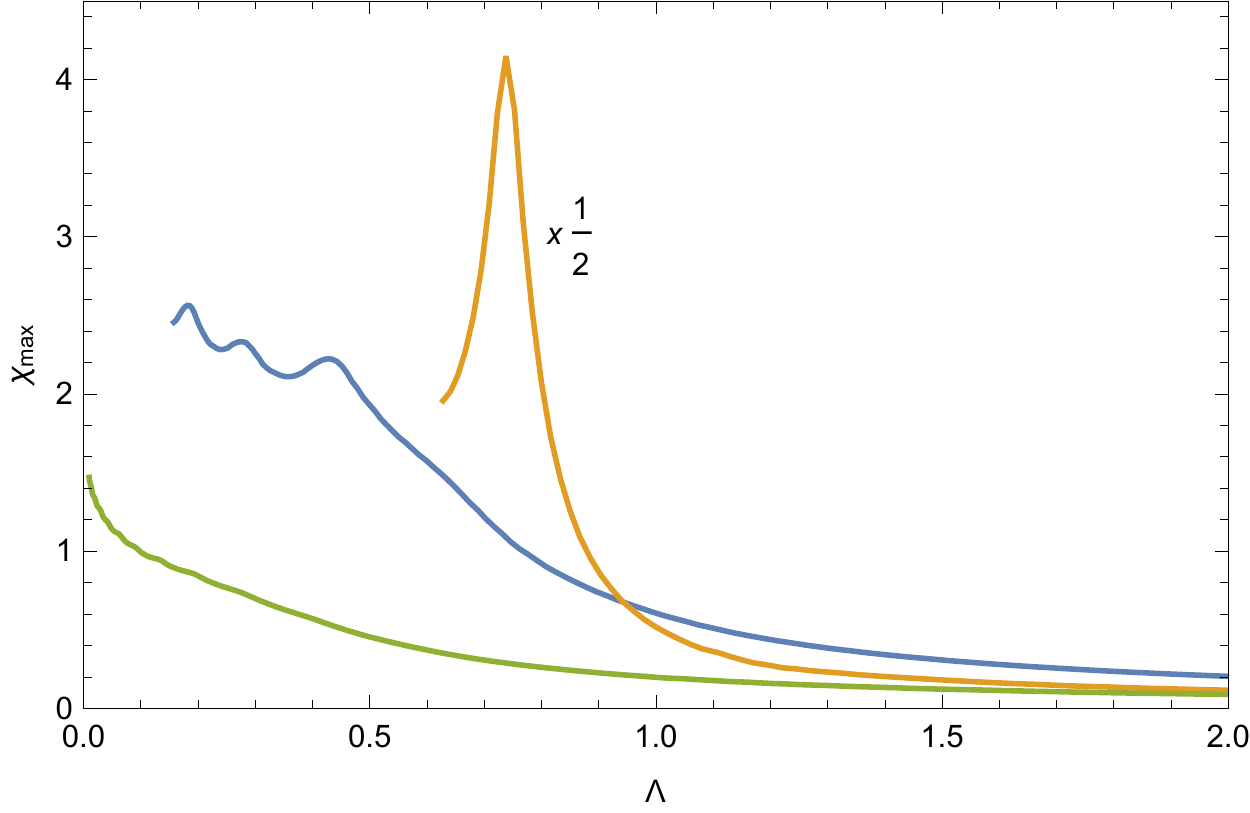}
    \caption{Maximal susceptibility in momentum space as a function of the renormalization group parameter $\Lambda$ from PFFRG. The green curve is an example for a non-magnetic system (spin-$1/2$ nearest neighbor antiferromagnetic Heisenberg model on the pyrochlore lattice). The orange curve represents a magnetically ordered system (spin-1 Heisenberg model on the double trillium lattice with $J_4>0$ couplings only). The blue curve represents the data for {\kni}.}
    \label{fig:PFFRGlambda}
\end{figure}

\end{document}